\documentclass[twocolumn]{aastex62}

\newcommand*\inst[1]{\unskip\hbox{\@textsuperscript{\normalfont$#1$}}}

\newcommand*\institute[1]{
  \begingroup
    \let\and\relax
    \renewcommand*\inst[1]{}%
    \renewcommand*\thanks[1]{}%
    \renewcommand*\email[1]{}%
  \endgroup
  \newcommand{\institutions}{#1}
}%

\let\oldarcsec\arcsec
\renewcommand\arcsec{\oldarcsec\xspace}%

\usepackage{latexsym}		
\usepackage{graphicx}		
\usepackage{rotating}		
\usepackage{natbib}  
\usepackage{savesym}
\usepackage{amssymb}
\usepackage{amsmath}
\usepackage{morefloats}
\savesymbol{doublespace}
\usepackage{xspace}
\usepackage{color}
\usepackage{mdframed}
\usepackage{url}
\usepackage{subfigure}
\usepackage{grffile}
\usepackage{import}
\usepackage[utf8]{inputenc}
\usepackage{booktabs}
\usepackage{fancyhdr}

\usepackage[hang,flushmargin]{footmisc}
\usepackage{ifpdf}

\newcommand{\msun}{\ensuremath{M_{\odot}}\xspace}			

\newcommand{\lsun}{\ensuremath{L_{\odot}}\xspace}			
\newcommand{\rsun}{\ensuremath{R_{\odot}}\xspace}			
\newcommand{\hh}{\ensuremath{\textrm{H}_{2}}\xspace}			

\newcommand{\water}{H$_{2}$O\xspace}		

\newcommand{\kms}{\textrm{km~s}\ensuremath{^{-1}}\xspace}	

\newcommand{\pers}{\ensuremath{\mathrm{s}^{-1}}\xspace}

\newcommand{\percc}{\ensuremath{\textrm{cm}^{-3}}\xspace}

\newcommand{\um}{\ensuremath{\mu \textrm{m}}\xspace}    

\def\ee#1{\ensuremath{\times10^{#1}}}
\newcommand{\degrees}{\ensuremath{^{\circ}}}

\newcommand{\perbeam}{\ensuremath{\textrm{beam}^{-1}}\xspace}

\def\eqref#1{Equation \ref{#1}}

\newenvironment{rotatepage}
{}{}

\newcommand{\githash}{5636c86}\newcommand{\gitdate}{2019-01-01\xspace}\newcommand{\sourcei}{SrcI\xspace}

\newcommand{\referee}[1]{#1}
\begin{document}
\newcommand{\nraojansky}{\affiliation{\it{Jansky fellow of the National Radio Astronomy Observatory, 1003 Lopezville Rd, Socorro, NM 87801 USA }}}
\newcommand{\nrao}{\affiliation{\it{National Radio Astronomy Observatory, 1003 Lopezville Rd, Socorro, NM 87801 USA }}}
\newcommand{\nraocv}{\affiliation{\it{National Radio Astronomy Observatory, Charlottesville, VA 22903 USA }}}
\newcommand{\cfa}{\affiliation{\it{Harvard-Smithsonian Center for Astrophysics, Cambridge, MA 02138 USA }}}
\newcommand{\hubble}{\altaffiliation{B.A.M. is a Hubble Fellow of the National Radio Astronomy\\ Observatory.}}

\newcommand{\radboud}{\affiliation{\it{Department of Astrophysics/IMAPP, Radboud University Nijmegen, PO Box 9010, 6500 GL Nijmegen, the Netherlands}}}
\newcommand{\allegro}{\affiliation{\it{ALLEGRO/Leiden Observatory, Leiden University, PO Box 9513, 2300 RA Leiden, the Netherlands}}}
\newcommand{\casa}{\affiliation{\it{CASA, University of Colorado, 389-UCB, Boulder, CO 80309}} }

\newcommand{\berkeley}{\affiliation{\it{Radio Astronomy Laboratory, University of California, Berkeley, CA 94720}} }

\author[0000-0001-6431-9633]{Adam Ginsburg}
\nraojansky

\correspondingauthor{Adam Ginsburg}
\email{aginsbur@nrao.edu; adam.g.ginsburg@gmail.com}

\author{Brett McGuire}
\hubble
\nraocv
\cfa

\author{Richard Plambeck}
\berkeley

\author{John Bally}
\casa

\author{Ciriaco Goddi}
\allegro
\radboud

\author{Melvyn Wright}
\berkeley

\title{Orion \sourcei's disk is salty}

\begin{abstract}
    We report the detection of NaCl, KCl, and their $^{37}$Cl and $^{41}$K
    isotopologues toward the disk around Orion \sourcei.  About 60 transitions
    of these molecules were identified.
    This is the first detection of these molecules in the interstellar
    medium not associated with the ejecta of evolved stars.  It is also
    the first ever detection of the vibrationally excited states of these
    lines in the ISM above $v=1$, with firm detections up to $v=6$.
    The salt emission traces the region just above the continuum disk,
    possibly forming the base of the outflow.  The emission from the
    vibrationally excited transitions is inconsistent with a single
    temperature, implying the lines are not in LTE.  We examine several
    possible explanations of the observed high excitation lines, concluding
    that the vibrational states are most likely to be radiatively excited via
    rovibrational transitions in the 25-35 \um (NaCl) and 35-45 \um (KCl)
    range.  We suggest that the molecules are produced by destruction of dust
    particles.  Because these molecules are so rare, they are potentially
    unique tools for identifying high-mass protostellar disks and measuring the
    radiation environment around accreting young stars.\\
\end{abstract}

\section{Introduction}
The disk around Orion Source I (\sourcei) has been the subject of intense
study, as it is the closest \citep[$d\approx400$ pc;][]{Grossschedl2018a} disk
around a `high-mass' ($M>8$ \msun) star
\citep{Hirota2014a,Plambeck2016a,Ginsburg2018b}.  It is also in many ways
a unique source, being the only known protostellar object with \referee{both}
SiO and \water
masers in an outflow
\citep{Plambeck2009a,Goddi2009a,Matthews2010a,Goddi2010a,Niederhofer2012a,Greenhill2013a}.
In \citet{Ginsburg2018b}, a subset of the authors reported 0.03$\arcsec$ ALMA
observations of \sourcei that allowed
a measurement of the outer rotation curve of
the 5(5,0)-6(4,3) \water line in the disk and along the base of the outflow;
these measurements showed that the interior mass was $15 \pm 2$ \msun.  In the
same data, we found additional emission lines that closely traced the disk and
its Keplerian rotation curve, but were unable to identify the molecular species
responsible for these lines.  We now identify them as transitions of NaCl, 
KCl, and their isotopologues, many in excited vibrational levels.

Disks are ubiquitous around forming stars of any mass.  They provide the main
mechanism for mediating accretion, and, via outflows, shedding angular momentum
from the system.  Despite their theoretical importance, the role of disks in
high-mass star formation remains observationally uncertain, since the only
definitive disk detections are around stars that have already acquired most of
their mass \citep[e.g.,][]{Girart2017a}.  The lack of clear disk detections
is, at least in part, because there were no known spectral lines that trace a
disk and not the molecule-rich ``hot core'' in the surrounding region
\citep{Goddi2018a,Cesaroni2017a}.  For low-mass sources, where the disks are
observed in isolation after the surrounding core has accreted or dispersed,
this ambiguity is absent.  For high-mass sources, which evolve so quickly that
the disks are still embedded in their natal hot core, substantial confusion can
arise.  Further complicating matters, the large amount of dust in these dense
regions may obscure the disks at wavelengths shorter than 3 mm.

While dozens of species have been detected in the disks around low-mass stars
\citep{McGuire2018c}, these molecules are comprised solely of a small selection
of elements: H, C, N, O, and S.  All of the molecules seen in disks are also
prevalent in the ISM, particularly in the hot cores surrounding high-mass stars
\citep{Nummelin1998a,Belloche2013a}. In high-mass environments, this therefore
limits their utility as unique disk tracers, hiding the kinematic signatures of
rotation.

Molecules consisting of alkali metals and halogens have only rarely
been detected in space \citep{McGuire2018c}.  Because of their rarity, their
use as a diagnostic tool to measure metallicity and local physical conditions
has been limited.  However, some of these molecules, such as KCl and NaCl, have
a rich spectrum with dozens of lines arising in a single observing band
so they have great potential to probe either local gas
properties or the radiation field when they are detected.

KCl and NaCl have so far been observed toward evolved stars including the
carbon star IRC +10216 \citep{Cernicharo1987a}, the oxygen-rich evolved stars IK
Tauri and VY Canis
Majoris \citep{Milam2007a}, the post-AGB star envelope CRL 2688
\citep{Highberger2003a}, and the AGB star wind around OH 231+4.2
\citep{Sanchez-Contreras2018a}.  In these sources, they exist only in a limited
range of radii from the central star as they are
carried outward
in slow winds \citep{Herwig2005a}.  The limited environments in which these
molecules have been detected suggests that they persist in the gas phase
for only a brief period 
before they are incorporated into dust grains.
This transient gas-phase abundance is analogous to that which makes SiO an
excellent tracer of recent shock events: Si is sputtered from grains into the
gas phase where it reacts with oxygen to form SiO.  Further reactions rapidly
form SiO$_2$, depleting the gas-phase SiO abundance, and making that molecule
an indicator of a recent shock \citep{Schilke1997a}. 

If the production pathways, and role of the physical environment on those
pathways, can be constrained for NaCl and KCl, the presence of these molecules
has the potential to be a powerful tracer of gas physical conditions and
history.  The rarity of these molecules in the ISM makes them a uniquely
powerful tracer when they can be found.  Below, we describe the first
detection of salts in a high-mass protostellar disk, unveiling what may
be the only molecules that specifically trace disks around high-mass stars.

\section{Observations and Analysis}
\label{sec:observations}

The observations presented here are described in \citet{Ginsburg2018b} as part
of ALMA project 2016.1.00165.S.  We use the robust 0.5 weighted spectral cubes
from all three bands (B3 3.0 mm, B6 1.3 mm, B7 0.87 mm) for our spectroscopic analysis.

\begin{figure*}[!htp]
\includegraphics[scale=1,width=5.5in]{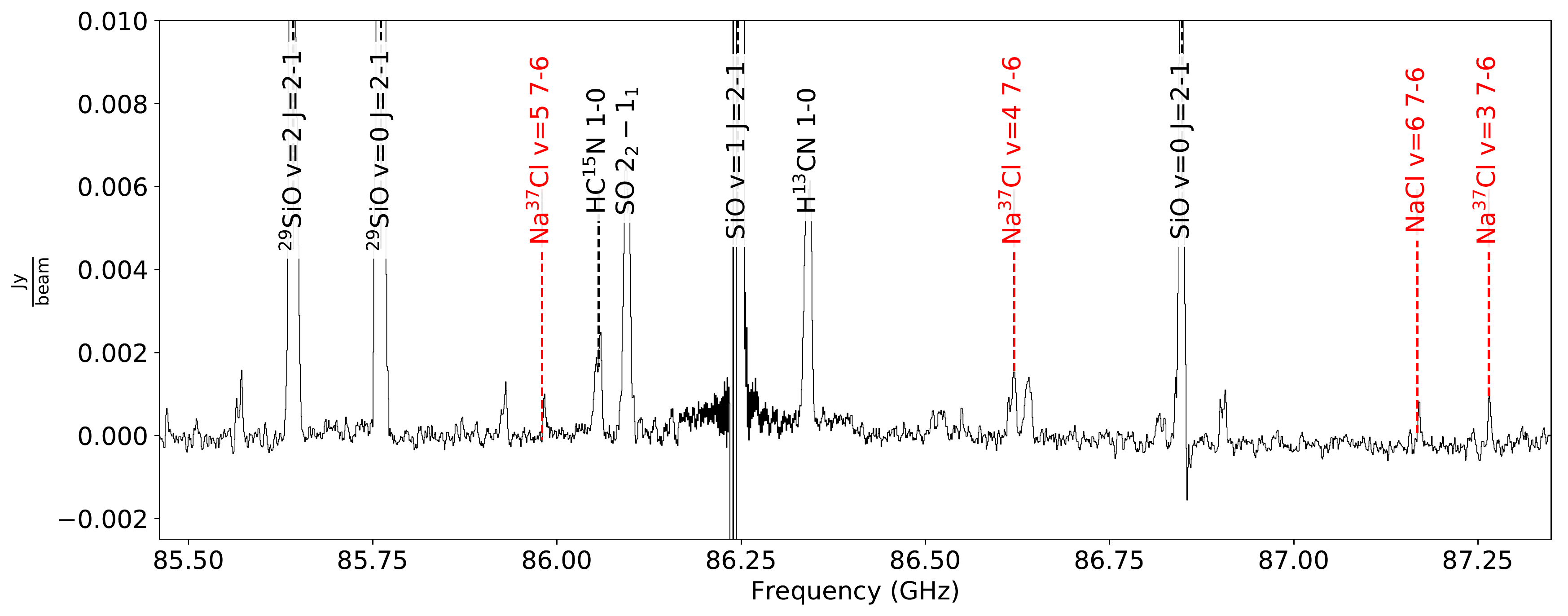}
\includegraphics[scale=1,width=5.5in]{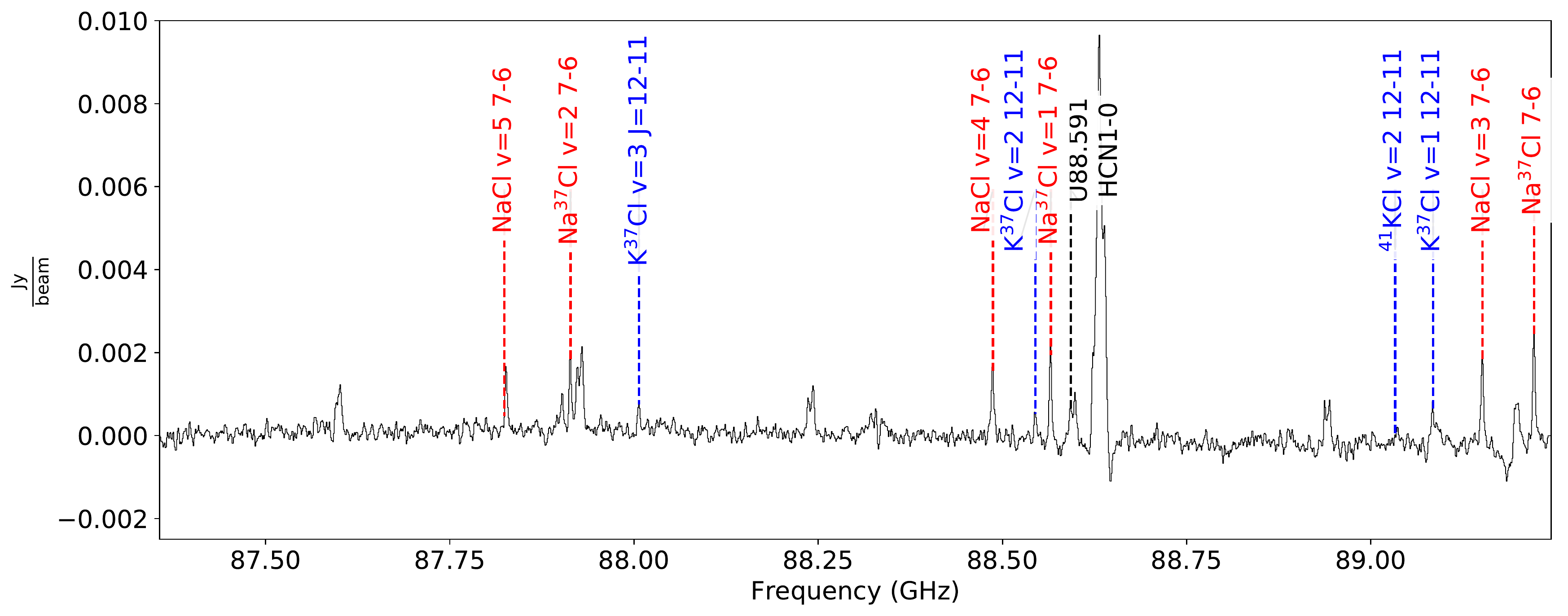}
\includegraphics[scale=1,width=5.5in]{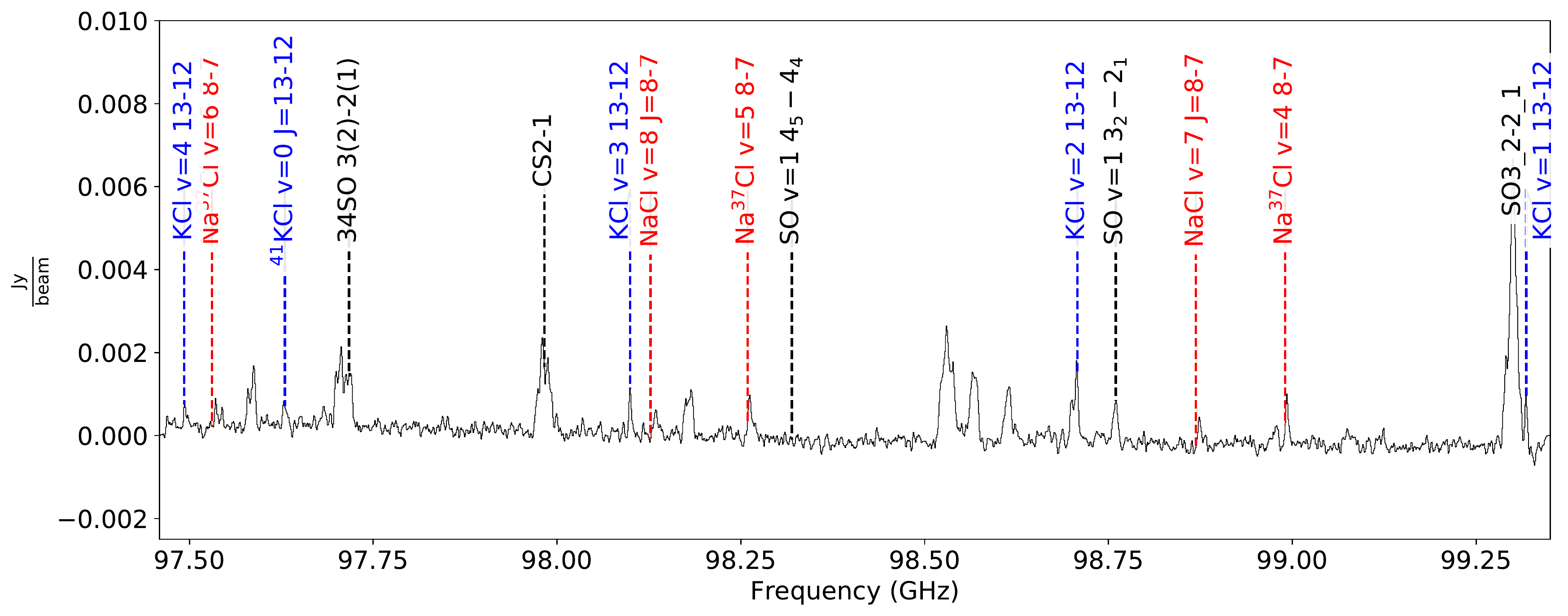}
\includegraphics[scale=1,width=5.5in]{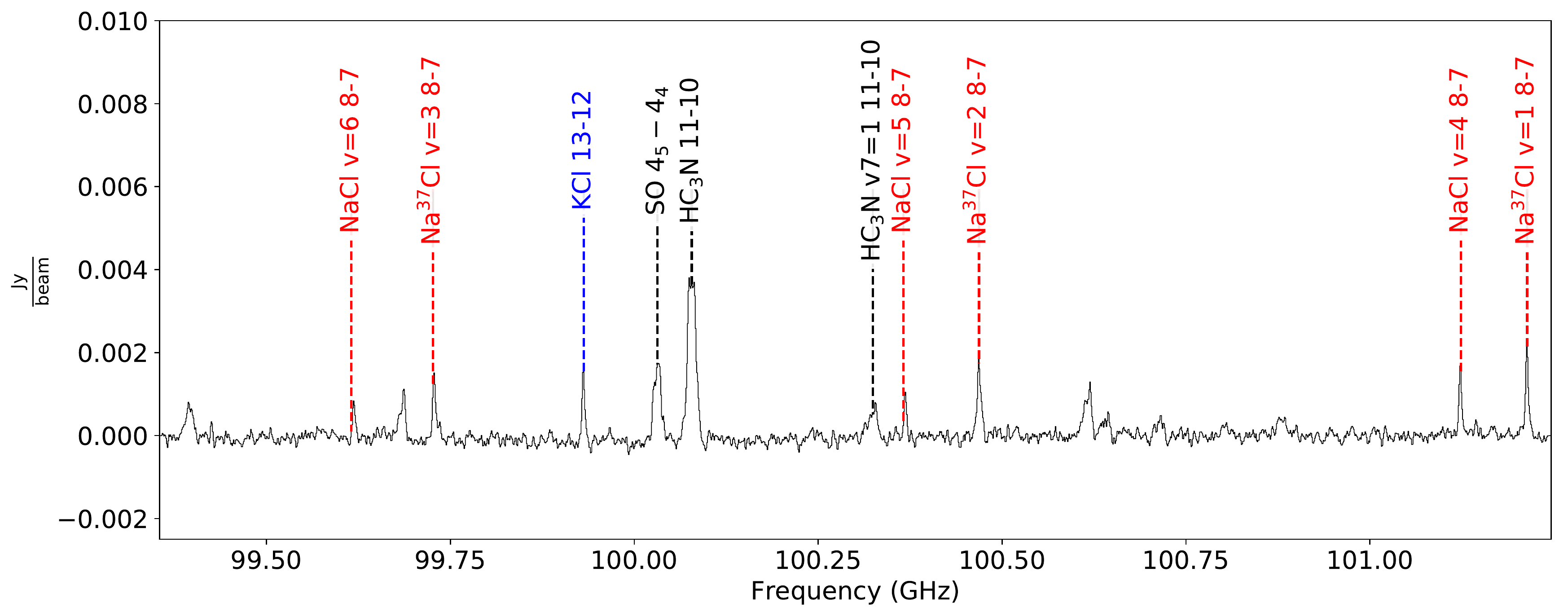}
\caption{Stacked spectra from the Band 3 setup (see Section \ref{sec:observations}).}
\label{fig:spectrab3}
\end{figure*}

\begin{figure*}[!htp]
\includegraphics[scale=1,width=5.5in]{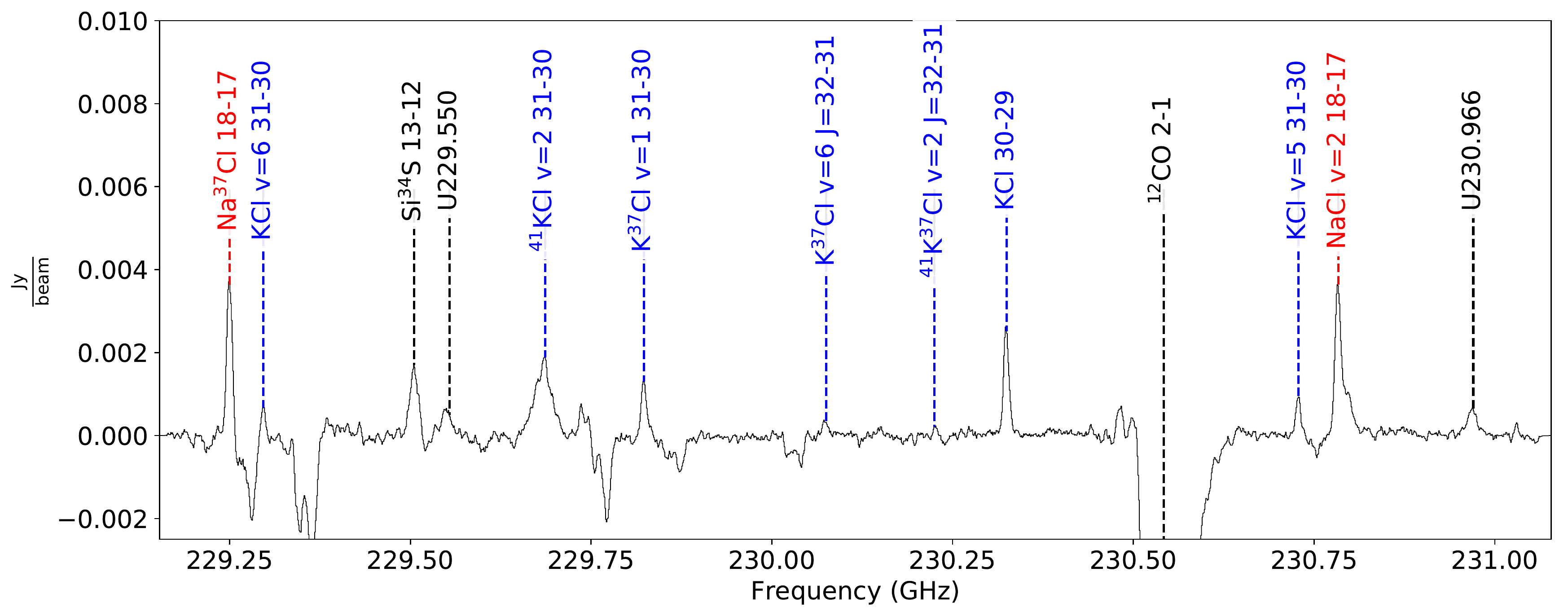}
\includegraphics[scale=1,width=5.5in]{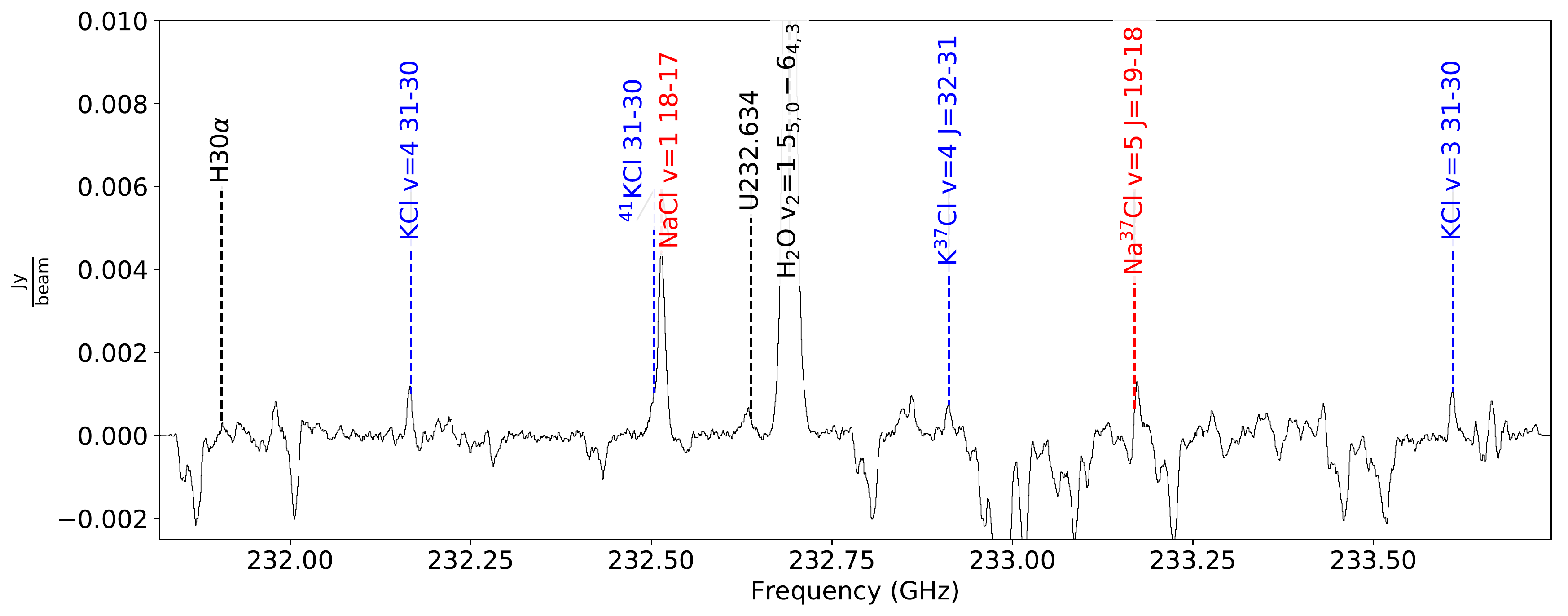}
\includegraphics[scale=1,width=5.5in]{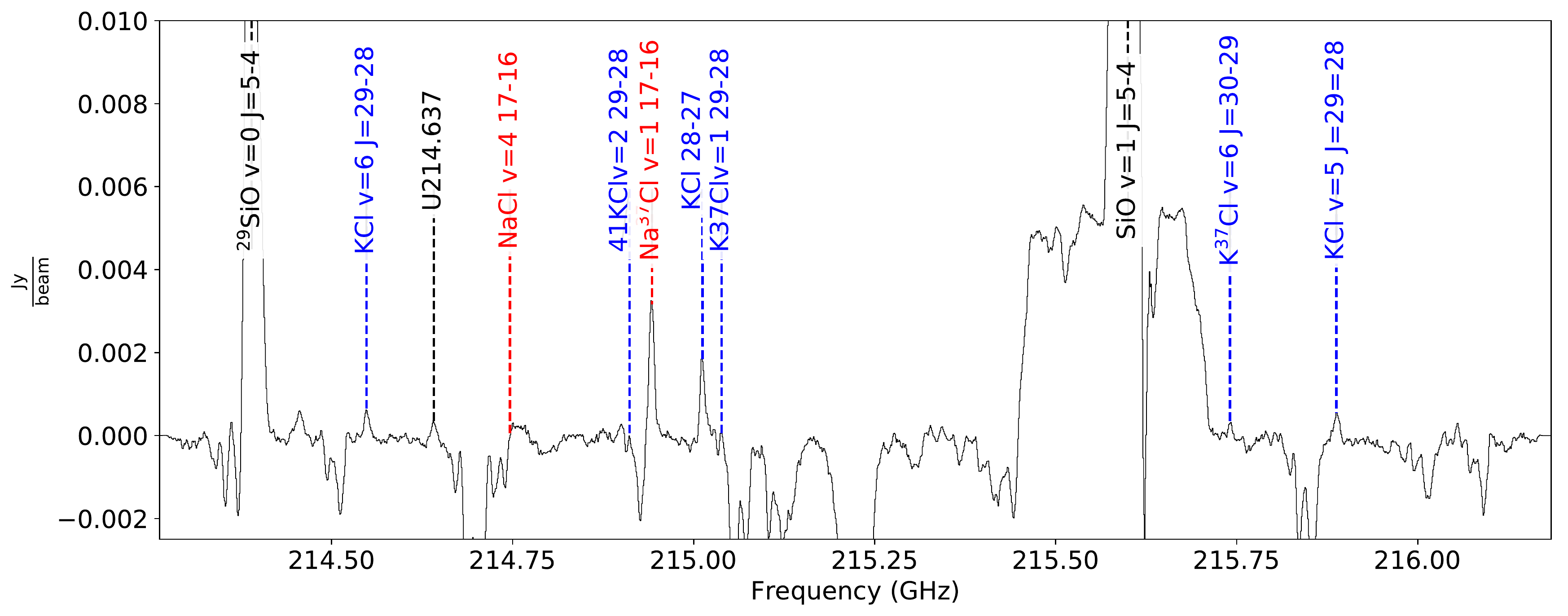}
\includegraphics[scale=1,width=5.5in]{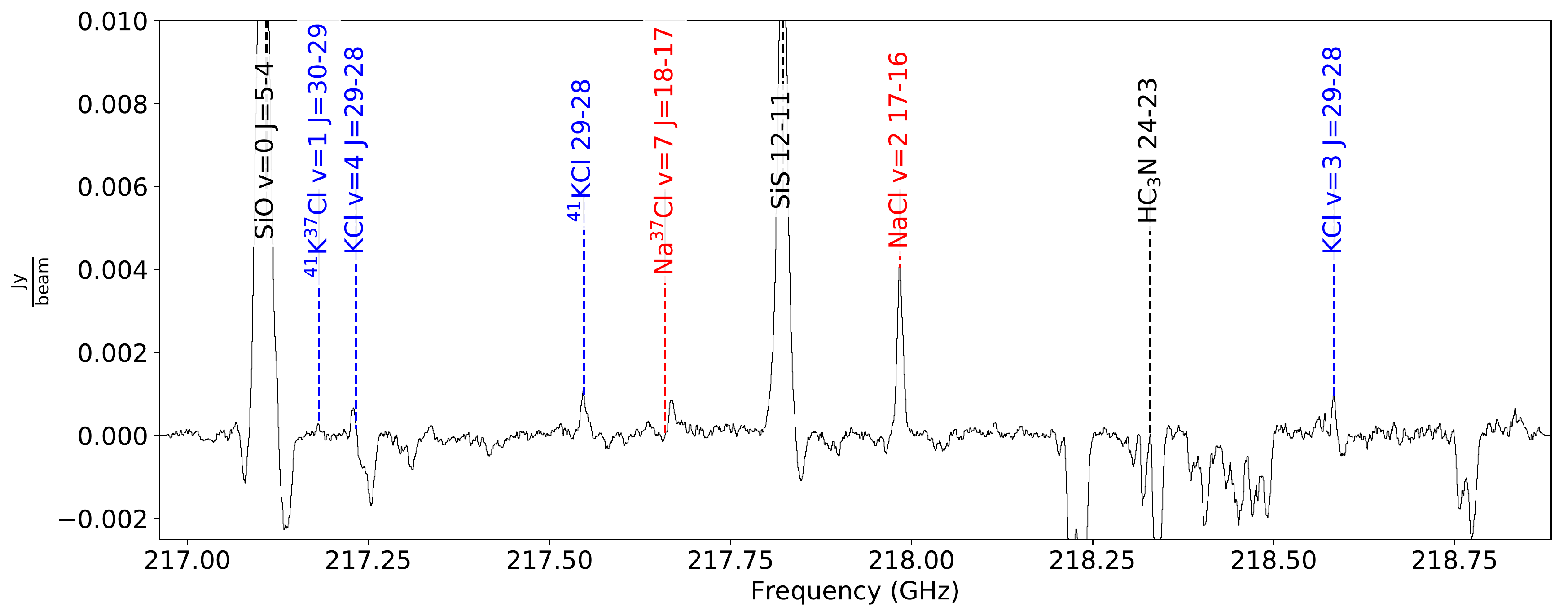}
\caption{Stacked spectra from the Band 6 setup (see Section \ref{sec:observations}).}
\label{fig:spectrab6}
\end{figure*}

\begin{figure*}[!htp]
\includegraphics[scale=1,width=5.5in]{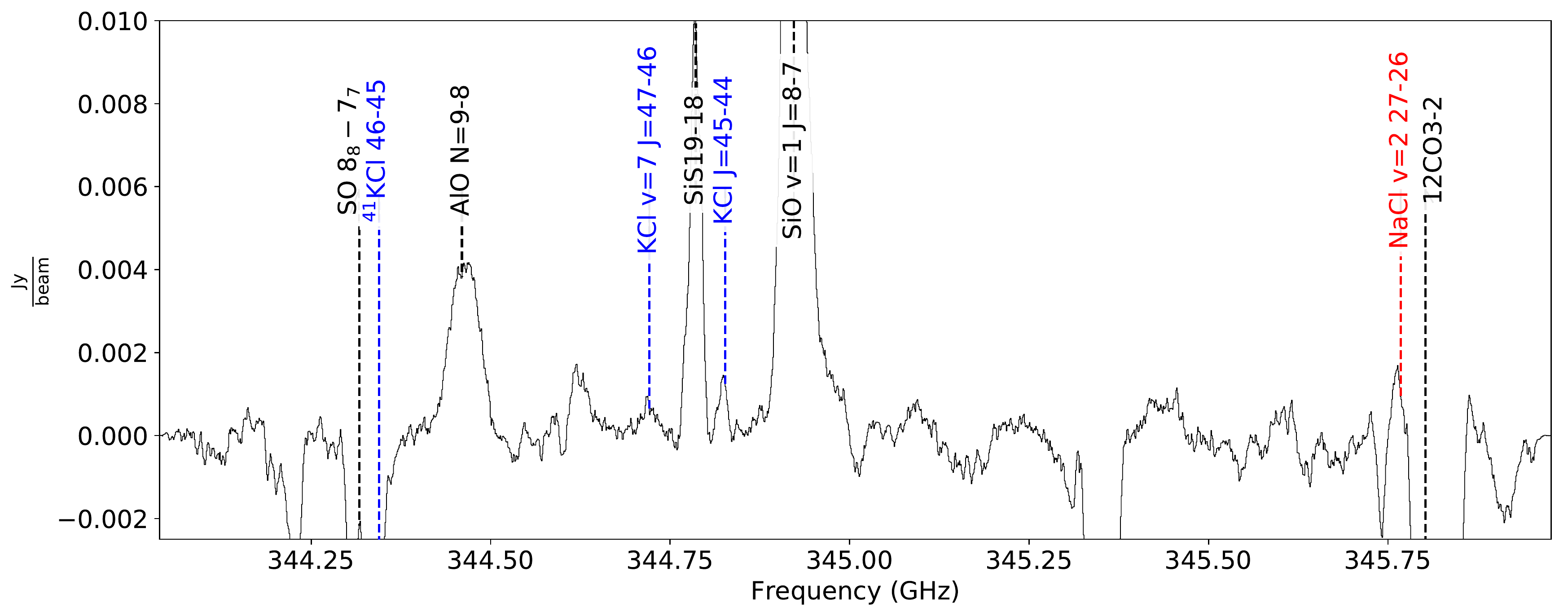}
\includegraphics[scale=1,width=5.5in]{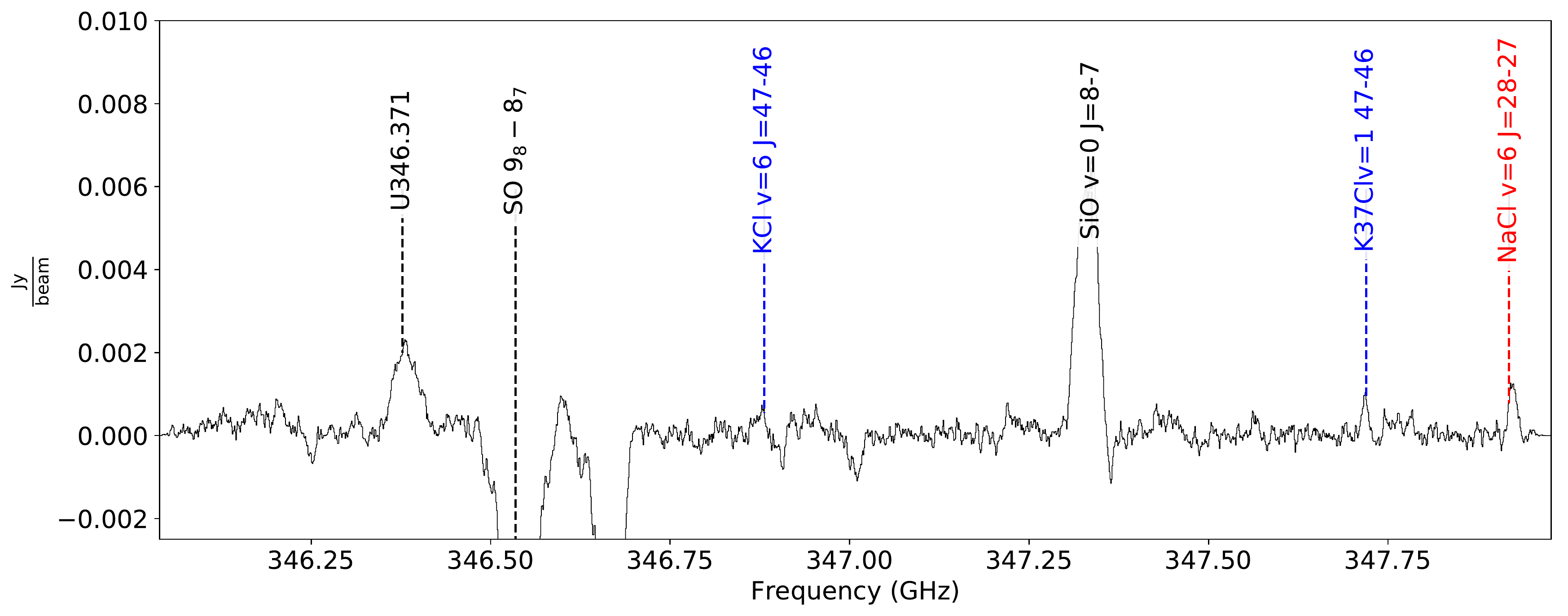}
\includegraphics[scale=1,width=5.5in]{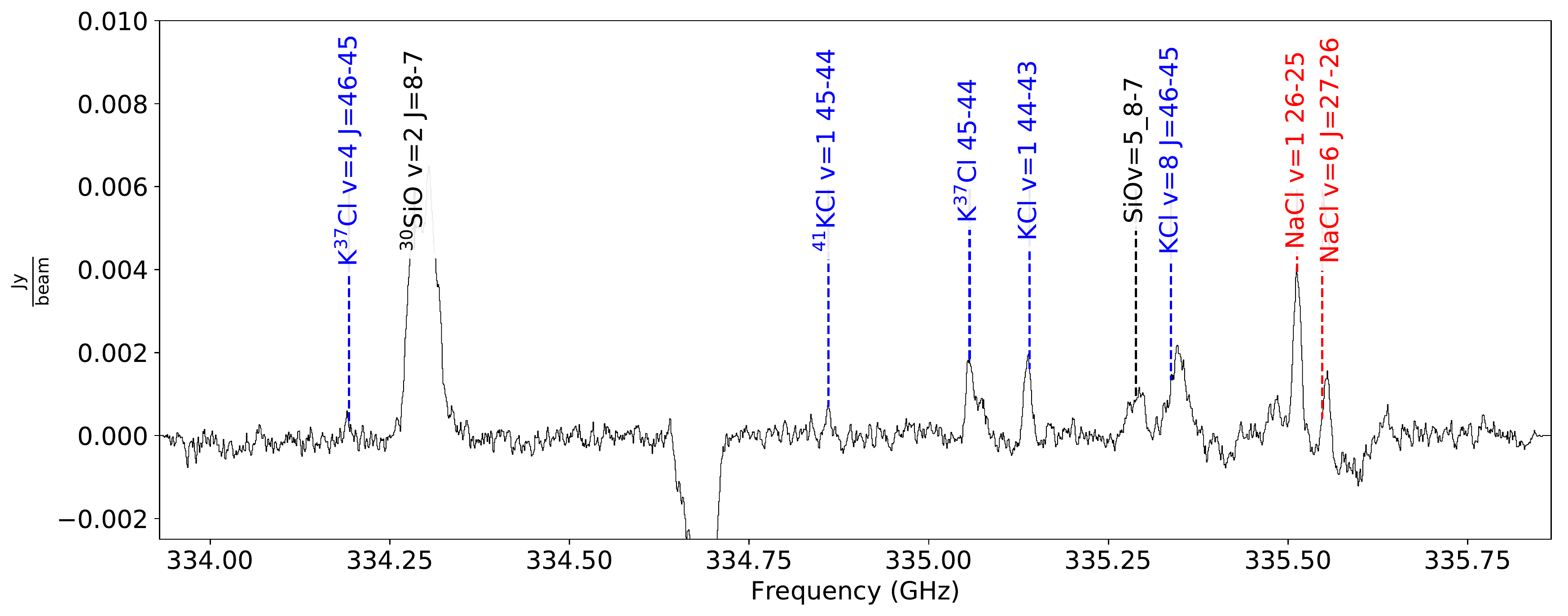}
\includegraphics[scale=1,width=5.5in]{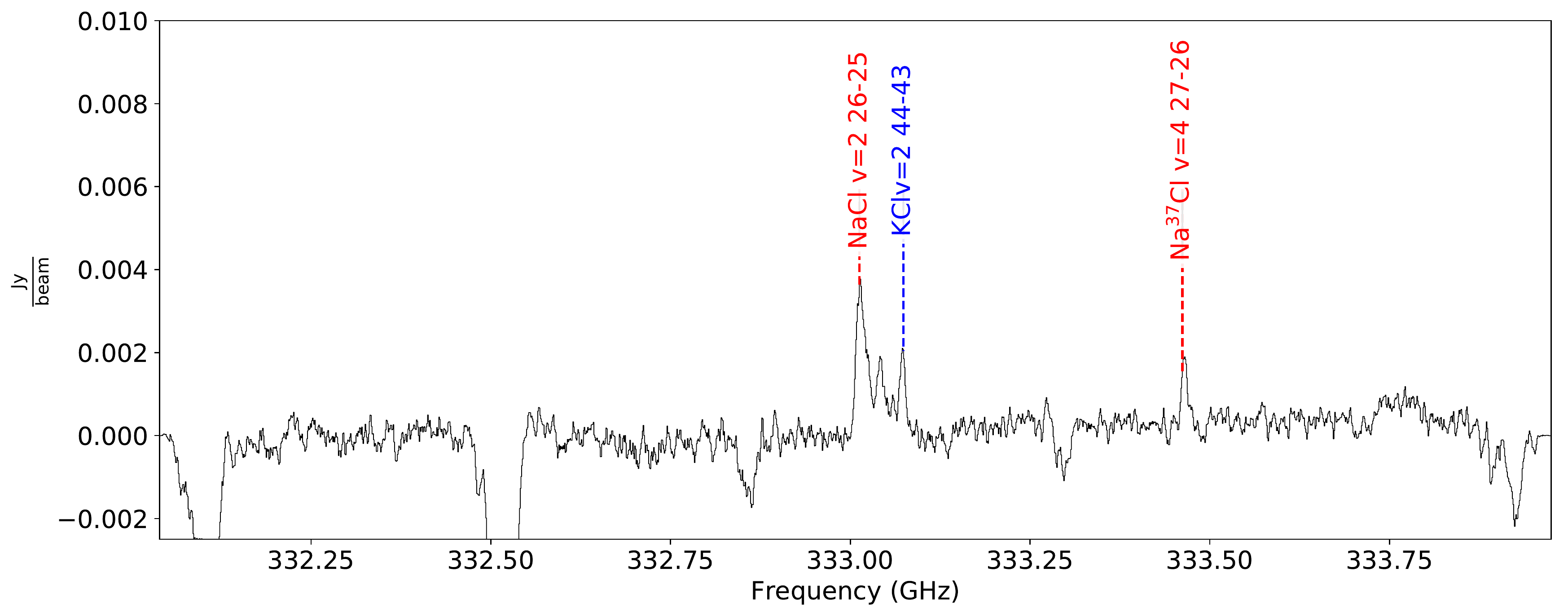}
\caption{Stacked spectra from the Band 7 setup (see Section \ref{sec:observations}).}
\label{fig:spectrab7}
\end{figure*}

Appendix D of that paper describes the spectral extraction method, which we
summarize here.  We used the U232.511 line (which we now identify as NaCl v=1
J=18-17) to find the velocity centroid for each (spatial) pixel across the
\sourcei disk.  We shifted the spectrum along each such pixel to 0 \kms, then
averaged the shifted spectra over the region with significant emission in the
NaCl line.  The averaging area is approximately the extent of the continuum
disk, 0.03 square arcseconds, or about 4, 20, and 40  beam areas, resulting in
an improvement in the signal-to-noise of about $2\times$, $4\times$, and
$6\times$ respectively in Band 3, 6, and 7.  This stacked spectrum, shown in
Figures \ref{fig:spectrab3}-\ref{fig:spectrab7}, traces material just above and
below the optically thick continuum disk (Figure \ref{fig:spatial}).  Because
this procedure removes the rotational velocity field of the disk, lines in the
stacked spectrum are narrower than one would observe in a lower spatial
resolution observation of the disk.  One may think of the stacked spectrum as
the output of a matched filter that is optimized for detection of disk
emission.

In the appendix of \citet{Ginsburg2018b}, we listed over 20 unidentified 
lines in the B6 ALMA spectra, and commented that ``there is no
consistent pattern to the detected lines and no individual species can explain
more than a few of the observed lines.''  This statement was incorrect, as
there are obvious carriers for the majority of the unidentified lines that we
had simply overlooked: NaCl, KCl, and their isotopologues.  The
detected lines have amplitudes in the range 0.5-3 mJy \perbeam, corresponding
to brightness temperatures 5-20 K.  

In Figures \ref{fig:spectrab3}-\ref{fig:spectrab7}, we have labeled all of the
detections and marginal detections of salt lines, as well as the most prominent
outflow (e.g., SiO and H$_2$O) lines.  Only 2-3 emission lines are now
unidentified in the B6 and B7 spectra, though there are over a dozen in B3 that
we have not identified. 

We also tentatively identify broad lines at 229.7 GHz and 344.4 GHz as AlO.  If
the identification is correct, these lines are broad because of the rich
hyperfine spectrum of AlO.


In Tables \ref{tab:Na_detections_B3}-\ref{tab:K_detections_B7}, we list all
of the salt lines in vibrational energy levels up to v=8 that lie within our
observed bands and whether or not they were detected.  We employ the following
flag scheme: `D' for detection, `N' for nondetection, `Q' for questionable
(e.g., low signal-to-noise, but possibly detected, or in regions where the
noise is not Gaussian and may be affected by other lines).  We additionally
include a flag `C' for `confused', which we add to the flag string if the line
is blended with another brighter line of a different species.  There are robust
detections of about 60 lines, up to the v=6 vibrational level.

The majority of the NaCl and KCl lines that we detected are in excited
vibrational states.  Although transitions in higher vibrational states were
measured, and some reported, in \citet{Caris2004a}, they did not provide the
complete catalogs needed for this analysis, nor to our knowledge are these
catalogues available in any of the standard online databases \citep[i.e., CDMS,
SLAIM, JPL;][]{Muller2005a,Lovas2005b,Pickett1998a}.  We obtained the KCl
frequencies for these lines from the catalogs of \citet{Barton2014a}, which was
designed primarily for use in exoplanet atmospheres and includes over 10$^5$
transitions for each molecular species and isotopologue with levels up to
$E_U\sim5\ee{4}$ K, and the NaCl frequencies from \citet{Cabezas2016a}, who
provided more accurate potential energy and dipole moment functions. The
primary result of these catalogs for the purposes of this work is the
prediction of frequencies in vibrational states higher than those measured in
the laboratory.

In using these catalogs, we found a systematic discrepancy between the rest
frequencies reported by \citet{Barton2014a}, and those given by
\citet{Caris2004a} and \citet{Caris2002a}:
the KCl lines appear to be systematically offset by $\sim$15-20 \kms in the
Barton catalog.  The Barton data are also discrepant with the more recent
\citet{Cabezas2016a} work that cataloged only NaCl lines. The offset is a
function of the vibrational and rotational energy states, indicating that it
resulted from an incorrect rotational and/or distortion constant.  We correct
for this error by performing a bilinear fit in frequency as a function of $v_u$
and $J_u$.  In the transitions available in both catalogs the resulting
frequency differences are generally $<0.1$ \kms, which is more than sufficient
for a reliable match to our spectra.  We then applied these
fitted models to the higher-$v$ states tabulated by Barton to get corrected
rest frequencies for KCl, while for NaCl we used the \citet{Cabezas2016a}
values.  We note that the values printed in \citet{Barton2014a} differ from
those in the digital catalogs hosted on \url{exomol.com}, so we suspect the
error was merely a transcription error of some sort.

\begin{table*}[htp]
\centering
\caption{All cataloged NaCl lines in the Band 3 tuning}
\begin{tabular}{ccccccccc}
\label{tab:Na_detections_B3}
Na Isotope & Cl Isotope & v & J$_u$ & J$_l$ & E$_U$ & A$_{ul}$ & Frequency & Flag \\
\hline
23 & 35 & 8 & 7 & 6 & 4035.4 & 0.00031 & 85.87018 & N \\
23 & 37 & 5 & 7 & 6 & 2538.2 & 0.00030 & 85.98167 & D \\
23 & 35 & 7 & 7 & 6 & 3550.1 & 0.00031 & 86.51754 & Q \\
23 & 37 & 4 & 7 & 6 & 2043.6 & 0.00030 & 86.62109 & D \\
23 & 35 & 6 & 7 & 6 & 3060.0 & 0.00032 & 87.16921 & D \\
23 & 37 & 3 & 7 & 6 & 1544.3 & 0.00030 & 87.26464 & D \\
23 & 35 & 5 & 7 & 6 & 2565.2 & 0.00032 & 87.82519 & D \\
23 & 37 & 2 & 7 & 6 & 1040.1 & 0.00031 & 87.91232 & D \\
23 & 35 & 4 & 7 & 6 & 2065.5 & 0.00032 & 88.48549 & D \\
23 & 37 & 1 & 7 & 6 & 531.1 & 0.00031 & 88.56415 & D \\
23 & 35 & 3 & 7 & 6 & 1561.0 & 0.00033 & 89.15011 & D \\
23 & 37 & 0 & 7 & 6 & 17.1 & 0.00031 & 89.22011 & D \\
23 & 37 & 6 & 8 & 7 & 3032.7 & 0.00045 & 97.53449 & Q \\
23 & 35 & 8 & 8 & 7 & 4040.1 & 0.00047 & 98.13295 & Q \\
23 & 37 & 5 & 8 & 7 & 2542.9 & 0.00045 & 98.26052 & Q \\
23 & 35 & 7 & 8 & 7 & 3554.8 & 0.00047 & 98.87277 & Q \\
23 & 37 & 4 & 8 & 7 & 2048.4 & 0.00045 & 98.99127 & D \\
23 & 35 & 6 & 8 & 7 & 3064.8 & 0.00048 & 99.61753 & D \\
23 & 37 & 3 & 8 & 7 & 1549.1 & 0.00046 & 99.72675 & D \\
23 & 35 & 5 & 8 & 7 & 2570.0 & 0.00048 & 100.36722 & D \\
23 & 37 & 2 & 8 & 7 & 1045.0 & 0.00046 & 100.46695 & D \\
23 & 35 & 4 & 8 & 7 & 2070.4 & 0.00048 & 101.12183 & D \\
23 & 37 & 1 & 8 & 7 & 536.0 & 0.00047 & 101.21188 & D \\
\hline
\end{tabular}

\par 
\end{table*}
\begin{table*}[htp]
\centering
\caption{All cataloged NaCl lines in the Band 6 tuning}
\begin{tabular}{ccccccccc}
\label{tab:Na_detections_B6}
Na Isotope & Cl Isotope & v & J$_u$ & J$_l$ & E$_U$ & A$_{ul}$ & Frequency & Flag \\
\hline
23 & 35 & 4 & 17 & 16 & 2141.3 & 0.00480 & 214.74232 & CN \\
23 & 37 & 1 & 17 & 16 & 607.0 & 0.00460 & 214.93871 & D \\
23 & 37 & 8 & 18 & 17 & 4076.0 & 0.00520 & 216.05316 & N \\
23 & 37 & 7 & 18 & 17 & 3596.2 & 0.00520 & 217.66504 & Q \\
23 & 35 & 2 & 17 & 16 & 1128.4 & 0.00490 & 217.98023 & D \\
23 & 37 & 0 & 18 & 17 & 104.6 & 0.00550 & 229.24605 & D \\
23 & 37 & 7 & 19 & 18 & 3607.2 & 0.00610 & 229.73281 & CN \\
23 & 35 & 2 & 18 & 17 & 1139.5 & 0.00580 & 230.77917 & D \\
23 & 35 & 1 & 18 & 17 & 625.7 & 0.00580 & 232.50998 & D \\
23 & 35 & 8 & 19 & 18 & 4130.7 & 0.00650 & 232.85904 & Q \\
23 & 37 & 5 & 19 & 18 & 2633.6 & 0.00620 & 233.16920 & CD \\
\hline
\end{tabular}

\par 
\end{table*}
\begin{table*}[htp]
\centering
\caption{All cataloged NaCl lines in the Band 7 tuning}
\begin{tabular}{ccccccccc}
\label{tab:Na_detections_B7}
Na Isotope & Cl Isotope & v & J$_u$ & J$_l$ & E$_U$ & A$_{ul}$ & Frequency & Flag \\
\hline
23 & 35 & 2 & 26 & 25 & 1250.2 & 0.01800 & 333.00729 & D \\
23 & 35 & 7 & 27 & 26 & 3757.5 & 0.01900 & 333.03612 & CQ \\
23 & 37 & 4 & 27 & 26 & 2251.3 & 0.01800 & 333.45906 & D \\
23 & 35 & 1 & 26 & 25 & 737.2 & 0.01800 & 335.50656 & D \\
23 & 35 & 6 & 27 & 26 & 3269.0 & 0.01900 & 335.54793 & CQ \\
23 & 37 & 8 & 28 & 27 & 4211.3 & 0.02000 & 335.63074 & CN \\
23 & 35 & 7 & 28 & 27 & 3774.1 & 0.02100 & 345.31422 & CN \\
23 & 37 & 4 & 28 & 27 & 2267.9 & 0.02000 & 345.75480 & CN \\
23 & 35 & 2 & 27 & 26 & 1266.8 & 0.02000 & 345.76204 & CD \\
23 & 37 & 8 & 29 & 28 & 4228.0 & 0.02200 & 347.55955 & N \\
23 & 35 & 6 & 28 & 27 & 3285.7 & 0.02100 & 347.91891 & NC \\
\hline
\end{tabular}

\par 
\end{table*}
\begin{table*}[htp]
\centering
\caption{All cataloged KCl lines in the Band 3 tuning}
\begin{tabular}{ccccccccc}
\label{tab:K_detections_B3}
K Isotope & Cl Isotope & v & J$_u$ & J$_l$ & E$_U$ & A$_{ul}$ & Frequency & Flag \\
\hline
41 & 35 & 8 & 12 & 11 & 3092.4 & 0.00041 & 85.80766 & N \\
39 & 37 & 7 & 12 & 11 & 2713.3 & 0.00041 & 85.87379 & N \\
41 & 37 & 3 & 12 & 11 & 1183.9 & 0.00039 & 85.93776 & N \\
41 & 35 & 7 & 12 & 11 & 2720.7 & 0.00042 & 86.33986 & CN \\
39 & 37 & 6 & 12 & 11 & 2339.4 & 0.00041 & 86.40364 & N \\
41 & 37 & 2 & 12 & 11 & 801.6 & 0.00039 & 86.45668 & N \\
41 & 35 & 6 & 12 & 11 & 2345.7 & 0.00042 & 86.87413 & N \\
39 & 37 & 5 & 12 & 11 & 1962.2 & 0.00042 & 86.93553 & N \\
41 & 37 & 1 & 12 & 11 & 416.1 & 0.00040 & 86.97743 & Q \\
41 & 35 & 5 & 12 & 11 & 1967.6 & 0.00042 & 87.41041 & N \\
39 & 37 & 4 & 12 & 11 & 1581.9 & 0.00042 & 87.46938 & N \\
41 & 37 & 0 & 12 & 11 & 27.3 & 0.00040 & 87.50010 & Q \\
39 & 35 & 8 & 12 & 11 & 3127.7 & 0.00044 & 87.77742 & N \\
41 & 35 & 4 & 12 & 11 & 1586.3 & 0.00043 & 87.94866 & N \\
39 & 37 & 3 & 12 & 11 & 1198.3 & 0.00042 & 88.00523 & D \\
39 & 35 & 7 & 12 & 11 & 2751.8 & 0.00045 & 88.32895 & Q \\
41 & 35 & 3 & 12 & 11 & 1201.7 & 0.00043 & 88.48896 & CQ \\
39 & 37 & 2 & 12 & 11 & 811.5 & 0.00042 & 88.54307 & D \\
39 & 35 & 6 & 12 & 11 & 2372.8 & 0.00045 & 88.88253 & N \\
41 & 35 & 2 & 12 & 11 & 813.8 & 0.00043 & 89.03129 & D \\
39 & 37 & 1 & 12 & 11 & 421.4 & 0.00043 & 89.08292 & D \\
39 & 35 & 4 & 13 & 12 & 1609.5 & 0.00058 & 97.49133 & D \\
41 & 35 & 0 & 13 & 12 & 32.8 & 0.00056 & 97.62809 & D \\
41 & 37 & 7 & 14 & 13 & 2690.7 & 0.00061 & 97.85364 & N \\
39 & 35 & 3 & 13 & 12 & 1220.6 & 0.00059 & 98.09753 & D \\
41 & 37 & 6 & 14 & 13 & 2321.0 & 0.00061 & 98.44993 & N \\
39 & 35 & 2 & 13 & 12 & 828.3 & 0.00059 & 98.70595 & D \\
41 & 37 & 5 & 14 & 13 & 1948.2 & 0.00062 & 99.04845 & N \\
39 & 35 & 1 & 13 & 12 & 432.6 & 0.00059 & 99.31663 & D \\
39 & 37 & 8 & 14 & 13 & 3093.3 & 0.00065 & 99.56276 & N \\
41 & 37 & 4 & 14 & 13 & 1572.3 & 0.00062 & 99.64924 & N \\
39 & 35 & 0 & 13 & 12 & 33.6 & 0.00060 & 99.92952 & D \\
41 & 35 & 8 & 14 & 13 & 3101.7 & 0.00066 & 100.10138 & N \\
39 & 37 & 7 & 14 & 13 & 2722.6 & 0.00065 & 100.17822 & N \\
41 & 37 & 3 & 14 & 13 & 1193.2 & 0.00063 & 100.25229 & N \\
41 & 35 & 7 & 14 & 13 & 2730.0 & 0.00067 & 100.72190 & N \\
39 & 37 & 6 & 14 & 13 & 2348.7 & 0.00066 & 100.79603 & Q \\
41 & 37 & 2 & 14 & 13 & 810.9 & 0.00063 & 100.85758 & N \\
\hline
\end{tabular}

\par 
\end{table*}
\begin{table*}[htp]
\centering
\caption{All cataloged KCl lines in the Band 6 tuning}
\begin{tabular}{ccccccccc}
\label{tab:K_detections_B6}
K Isotope & Cl Isotope & v & J$_u$ & J$_l$ & E$_U$ & A$_{ul}$ & Frequency & Flag \\
\hline
39 & 37 & 7 & 30 & 29 & 2846.1 & 0.00654 & 214.41527 & N \\
39 & 35 & 6 & 29 & 28 & 2499.6 & 0.00648 & 214.54412 & D \\
41 & 37 & 3 & 30 & 29 & 1316.8 & 0.00626 & 214.57918 & N \\
41 & 35 & 2 & 29 & 28 & 940.8 & 0.00622 & 214.90740 & CQ \\
39 & 35 & 0 & 28 & 27 & 149.7 & 0.00609 & 215.00828 & D \\
39 & 37 & 1 & 29 & 28 & 548.5 & 0.00616 & 215.03464 & CQ \\
41 & 37 & 8 & 31 & 30 & 3187.6 & 0.00668 & 215.09368 & CN \\
41 & 35 & 7 & 30 & 29 & 2854.2 & 0.00665 & 215.57755 & CN \\
39 & 37 & 6 & 30 & 29 & 2473.0 & 0.00659 & 215.73679 & D \\
41 & 37 & 2 & 30 & 29 & 935.3 & 0.00631 & 215.87559 & N \\
39 & 35 & 5 & 29 & 28 & 2118.0 & 0.00653 & 215.88373 & D \\
39 & 37 & 5 & 30 & 29 & 2096.7 & 0.00664 & 217.06391 & Q \\
41 & 37 & 1 & 30 & 29 & 550.6 & 0.00635 & 217.17723 & Q \\
39 & 35 & 4 & 29 & 28 & 1733.2 & 0.00658 & 217.22891 & CD \\
41 & 35 & 0 & 29 & 28 & 156.7 & 0.00631 & 217.54317 & D \\
41 & 37 & 6 & 31 & 30 & 2452.9 & 0.00677 & 217.72366 & N \\
41 & 35 & 5 & 30 & 29 & 2102.8 & 0.00675 & 218.24805 & N \\
39 & 37 & 4 & 30 & 29 & 1717.2 & 0.00668 & 218.39657 & CN \\
41 & 37 & 0 & 30 & 29 & 162.6 & 0.00639 & 218.48414 & CN \\
39 & 35 & 3 & 29 & 28 & 1345.1 & 0.00662 & 218.57971 & D \\
39 & 35 & 6 & 31 & 30 & 2521.2 & 0.00793 & 229.29217 & CD \\
41 & 35 & 2 & 31 & 30 & 962.5 & 0.00761 & 229.68227 & CD \\
39 & 37 & 1 & 31 & 30 & 570.2 & 0.00753 & 229.81880 & D \\
41 & 35 & 7 & 32 & 31 & 2875.9 & 0.00808 & 229.90046 & Q \\
39 & 37 & 6 & 32 & 31 & 2494.7 & 0.00800 & 230.07072 & D \\
41 & 37 & 2 & 32 & 31 & 957.1 & 0.00766 & 230.22070 & Q \\
41 & 37 & 7 & 33 & 32 & 2843.6 & 0.00812 & 230.31852 & CN \\
39 & 35 & 0 & 30 & 29 & 171.4 & 0.00749 & 230.32064 & D \\
39 & 35 & 5 & 31 & 30 & 2139.8 & 0.00798 & 230.72399 & D \\
39 & 35 & 4 & 31 & 30 & 1755.1 & 0.00804 & 232.16185 & D \\
41 & 35 & 0 & 31 & 30 & 178.7 & 0.00771 & 232.49980 & CN \\
41 & 35 & 5 & 32 & 31 & 2124.8 & 0.00819 & 232.74872 & Q \\
39 & 37 & 4 & 32 & 31 & 1739.2 & 0.00812 & 232.90755 & D \\
41 & 37 & 0 & 32 & 31 & 184.6 & 0.00777 & 233.00319 & CN \\
41 & 37 & 5 & 33 & 32 & 2102.9 & 0.00823 & 233.12954 & CN \\
39 & 35 & 3 & 31 & 30 & 1367.1 & 0.00810 & 233.60570 & D \\
39 & 35 & 8 & 32 & 31 & 3285.5 & 0.00860 & 233.72540 & N \\
\hline
\end{tabular}

\par 
\end{table*}
\begin{table*}[htp]
\centering
\caption{All cataloged KCl lines in the Band 7 tuning}
\begin{tabular}{ccccccccc}
\label{tab:K_detections_B7}
K Isotope & Cl Isotope & v & J$_u$ & J$_l$ & E$_U$ & A$_{ul}$ & Frequency & Flag \\
\hline
39 & 37 & 5 & 46 & 45 & 2310.3 & 0.02398 & 332.14436 & CN \\
39 & 35 & 6 & 45 & 44 & 2712.4 & 0.02429 & 332.22193 & CQ \\
41 & 37 & 8 & 48 & 47 & 3413.8 & 0.02483 & 332.29761 & N \\
41 & 37 & 1 & 46 & 45 & 764.4 & 0.02295 & 332.34847 & N \\
41 & 35 & 2 & 45 & 44 & 1154.0 & 0.02331 & 332.81369 & N \\
39 & 37 & 1 & 45 & 44 & 761.8 & 0.02309 & 333.01833 & CN \\
39 & 35 & 2 & 44 & 43 & 1155.3 & 0.02336 & 333.06770 & D \\
39 & 37 & 8 & 47 & 46 & 3441.9 & 0.02504 & 333.10414 & N \\
41 & 37 & 4 & 47 & 46 & 1921.1 & 0.02398 & 333.43025 & N \\
41 & 35 & 5 & 46 & 45 & 2317.6 & 0.02438 & 333.95245 & N \\
39 & 37 & 4 & 46 & 45 & 1932.1 & 0.02415 & 334.18696 & N \\
39 & 35 & 5 & 45 & 44 & 2332.2 & 0.02446 & 334.29930 & D \\
41 & 37 & 7 & 48 & 47 & 3049.3 & 0.02501 & 334.32791 & CD \\
41 & 37 & 0 & 46 & 45 & 377.7 & 0.02311 & 334.35249 & CN \\
41 & 35 & 1 & 45 & 44 & 764.9 & 0.02347 & 334.85439 & D \\
41 & 35 & 8 & 47 & 46 & 3452.1 & 0.02545 & 334.89960 & N \\
39 & 37 & 0 & 45 & 44 & 370.4 & 0.02325 & 335.05072 & D \\
39 & 35 & 1 & 44 & 43 & 761.7 & 0.02353 & 335.13396 & D \\
39 & 37 & 7 & 47 & 46 & 3073.3 & 0.02522 & 335.16416 & Q \\
39 & 35 & 8 & 46 & 45 & 3479.1 & 0.02557 & 335.33093 & CQ \\
41 & 37 & 3 & 47 & 46 & 1544.2 & 0.02414 & 335.45237 & N \\
41 & 35 & 7 & 48 & 47 & 3099.1 & 0.02730 & 344.08981 & CN \\
41 & 35 & 0 & 46 & 45 & 389.0 & 0.02525 & 344.33834 & CN \\
39 & 37 & 6 & 48 & 47 & 2718.1 & 0.02705 & 344.35234 & CN \\
41 & 37 & 2 & 48 & 47 & 1180.5 & 0.02589 & 344.60998 & CQ \\
39 & 35 & 7 & 47 & 46 & 3122.1 & 0.02747 & 344.71476 & D \\
39 & 35 & 0 & 45 & 44 & 381.2 & 0.02534 & 344.82061 & D \\
41 & 35 & 3 & 47 & 46 & 1572.6 & 0.02637 & 345.37541 & CN \\
41 & 37 & 5 & 49 & 48 & 2327.8 & 0.02698 & 345.40918 & CN \\
39 & 37 & 2 & 47 & 46 & 1182.6 & 0.02612 & 345.59885 & CN \\
39 & 35 & 3 & 46 & 45 & 1578.4 & 0.02650 & 345.94809 & CN \\
41 & 35 & 6 & 48 & 47 & 2726.5 & 0.02750 & 346.22022 & N \\
39 & 37 & 5 & 48 & 47 & 2343.3 & 0.02724 & 346.47449 & N \\
41 & 37 & 1 & 48 & 47 & 797.3 & 0.02607 & 346.69248 & CN \\
39 & 35 & 6 & 47 & 46 & 2745.3 & 0.02767 & 346.87489 & D \\
39 & 37 & 8 & 49 & 48 & 3474.8 & 0.02837 & 347.16340 & N \\
41 & 35 & 2 & 47 & 46 & 1187.0 & 0.02655 & 347.49773 & Q \\
41 & 37 & 4 & 49 & 48 & 1954.1 & 0.02717 & 347.50839 & N \\
39 & 37 & 1 & 47 & 46 & 794.8 & 0.02630 & 347.71265 & D \\
\hline
\end{tabular}

\par 
\end{table*}

We have fit Gaussian line profiles to each of the transitions we flagged as
`detected' in Tables \ref{tab:Na_detections_B3}-\ref{tab:K_detections_B7}.
The fits are reported in Tables
\ref{tab:NaCl_salt_lines}-\ref{tab:41K37Cl_salt_lines}.  
A few of the integrated intensities (e.g., for the KCl v=5 45-44 line) are discrepantly
high \referee{compared to transitions at similar $v$ and $J$ states},
suggesting that they are blends with other species as denoted in Tables
\ref{tab:Na_detections_B3}-\ref{tab:K_detections_B7}.

\section{Discussion}
NaCl and KCl have only been seen in a 
handful of other sources, all of
which were moderately high-mass evolved stars blowing off their envelopes.  The
known detections include the carbon star envelope IRC +10216
\citep{Cernicharo1987a,Agundez2012a,Zack2011a}, the post-AGB star envelope CRL 2688
\citep{Highberger2003a}, and the oxygen-rich evolved star envelopes of VY Canis
Majoris and IK Tauri \citep{Milam2007a,Kaminski2013a}. To our knowledge,
\sourcei is only the
fifth astronomical source in which KCl and NaCl have been detected.
The only previous detections of vibrationally excited NaCl were 
in the v=1 level
in IRC+10216 \citep{Quintana-Lacaci2016a} and VY CMa \citep{Decin2016a}, 
while \sourcei exhibits clear emission up to v=6 in both NaCl and KCl.

In these evolved stars, v=0 rotational transitions of NaCl were observed
with rotational temperatures ${T_{rot}\sim70-100}$ K (IK Tau, VY CMa), inferred
local temperatures of $\sim200$ K in shocks (CRL 2688), and $T\approx700$ K  in
an expanding shell (IRC+10216).  Below, we attempt to determine the physical
conditions in the \sourcei salt emission regions.

\subsection{Spatial distribution of the salt emission}
\label{sec:wherefrom}
\begin{figure*}[!htp]
\includegraphics[scale=1,width=2.25in]{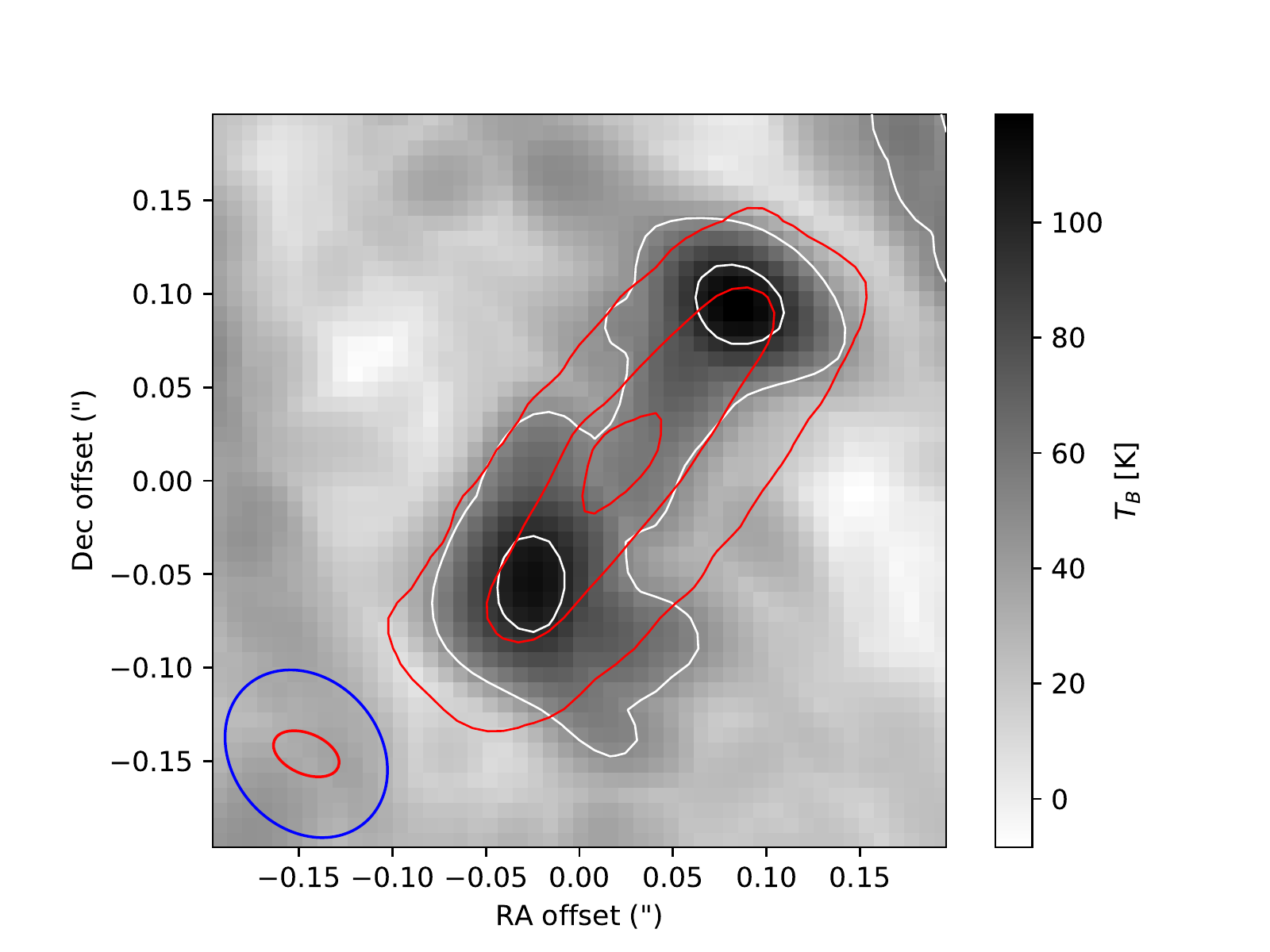}
\includegraphics[scale=1,width=2.25in]{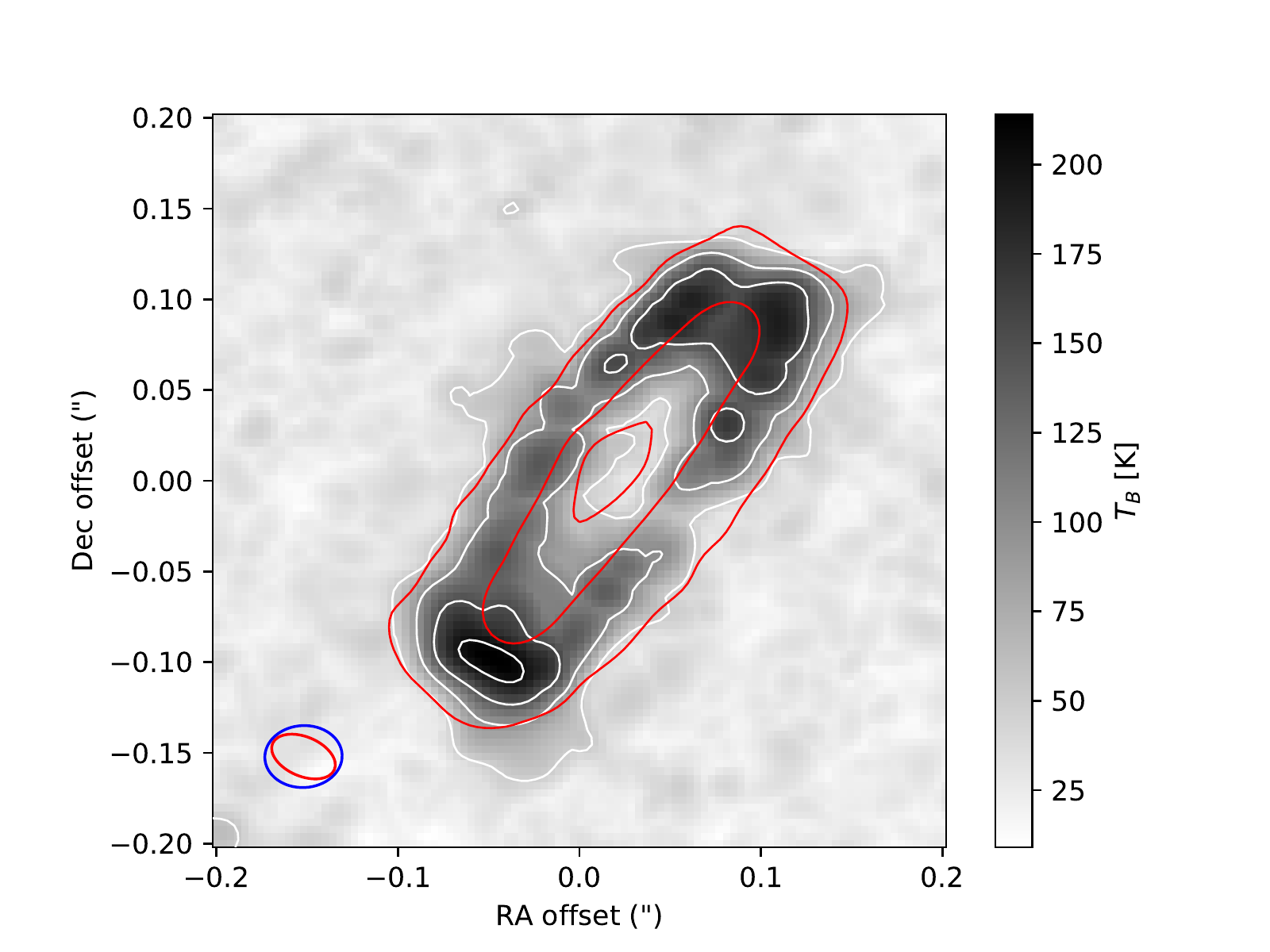}
\includegraphics[scale=1,width=2.25in]{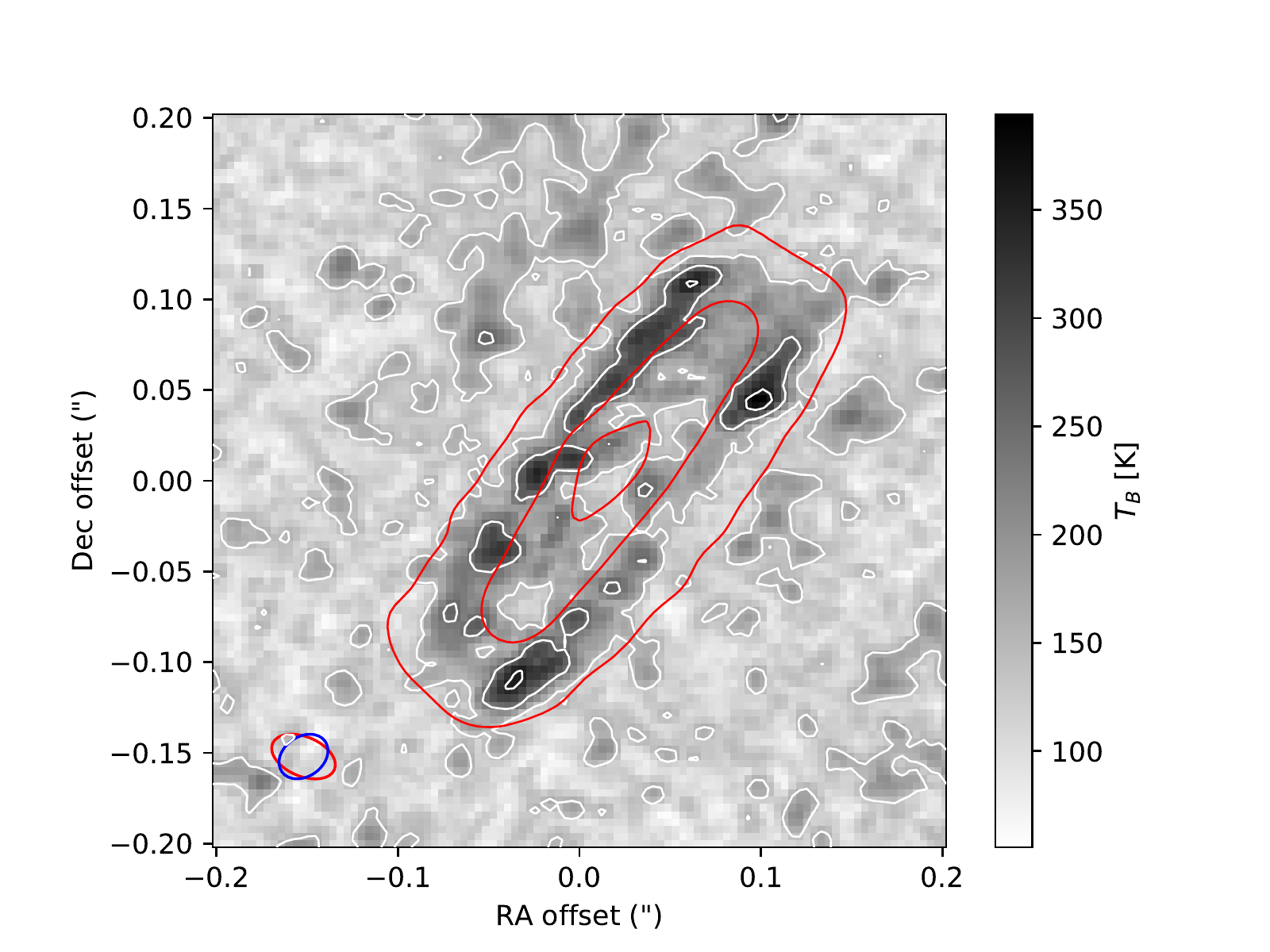}
\caption{Peak intensity images of NaCl lines in each band showing the spatial
distribution of the emission.  From left to right, the lines are: 87.26 GHz
NaCl v=3 J=7-6, 232.51 GHz NaCl v=1 J=18-17, and 333.01 GHz NaCl v=2 J=26-25.
These are the brightest uncontaminated lines in each of their observing bands.
The red contours show the Band 6 226 GHz continuum at levels of 50, 300, and 500 K.
White contours are shown at \referee{50 and 100 K (left), 50, 100,  150, and 200 K (center),
and 150, 250, and 350 K (right)}.
The red and blue ellipses show the beams for the continuum and the line emission,
respectively.  The beam sizes are
$0.10 \arcsec \times 0.08 \arcsec$, PA $40.6 \degrees$,
$0.043 \arcsec \times 0.034 \arcsec$, PA $-87.4 \degrees$, and
$0.029 \arcsec \times 0.022 \arcsec$, PA $-54.6 \degrees$, respectively.
}
\label{fig:spatial}
\end{figure*}

In \sourcei, the salt lines originate from the surface layers of the disk (the Band 7
lines exhibit NaCl emission peaks at $\pm0.032\arcsec=13$ AU above the disk),
unlike SiO and \water that both exhibit high vertical extents consistent with
outflow \citep{Ginsburg2018b}.  No emission in the salt lines is observed
toward the optically thick continuum in the disk midplane (Figure
\ref{fig:spatial}).  
\citet{Ginsburg2018b} modeled the line emission now assigned to these salts as a
truncated Keplerian disk, finding that the radial distribution of the emission
has an inner cutoff $\approx35-40$ AU and an outer cutoff $\approx55-60$~AU.

The equilibrium temperature in this radius range can be computed assuming the central
source has a luminosity $L=10^4 \lsun = 4 \pi r^2 \sigma_{SB} T^4$ \referee{($T_*=4000$ K, $R_*=210$ \rsun)}, where we
solve for temperature with $r$ at the inner and outer radius, giving $T_{eq} =
500-670$ K for $r=35-60$ AU, assuming the disk intercepts all of the starlight
at this radius.  More realistically, providing the opposite limiting case,
$T_{eq}=120-180$ K for a flat disk \citep[Eqn. 4 of][]{Chiang1997a}.  These
cooler temperatures are consistent with the observed brightness temperatures in
the outer part of the continuum disk (in Figure \ref{fig:spatial}, the \referee{Band 6
and 7} NaCl emission begins just beyond the $T=300$ K \referee{continuum}
contour), although the temperature in the inner portion of the disk reaches
$\gtrsim500$ K.

\begin{figure*}[!htp]
\includegraphics[scale=1,width=4in]{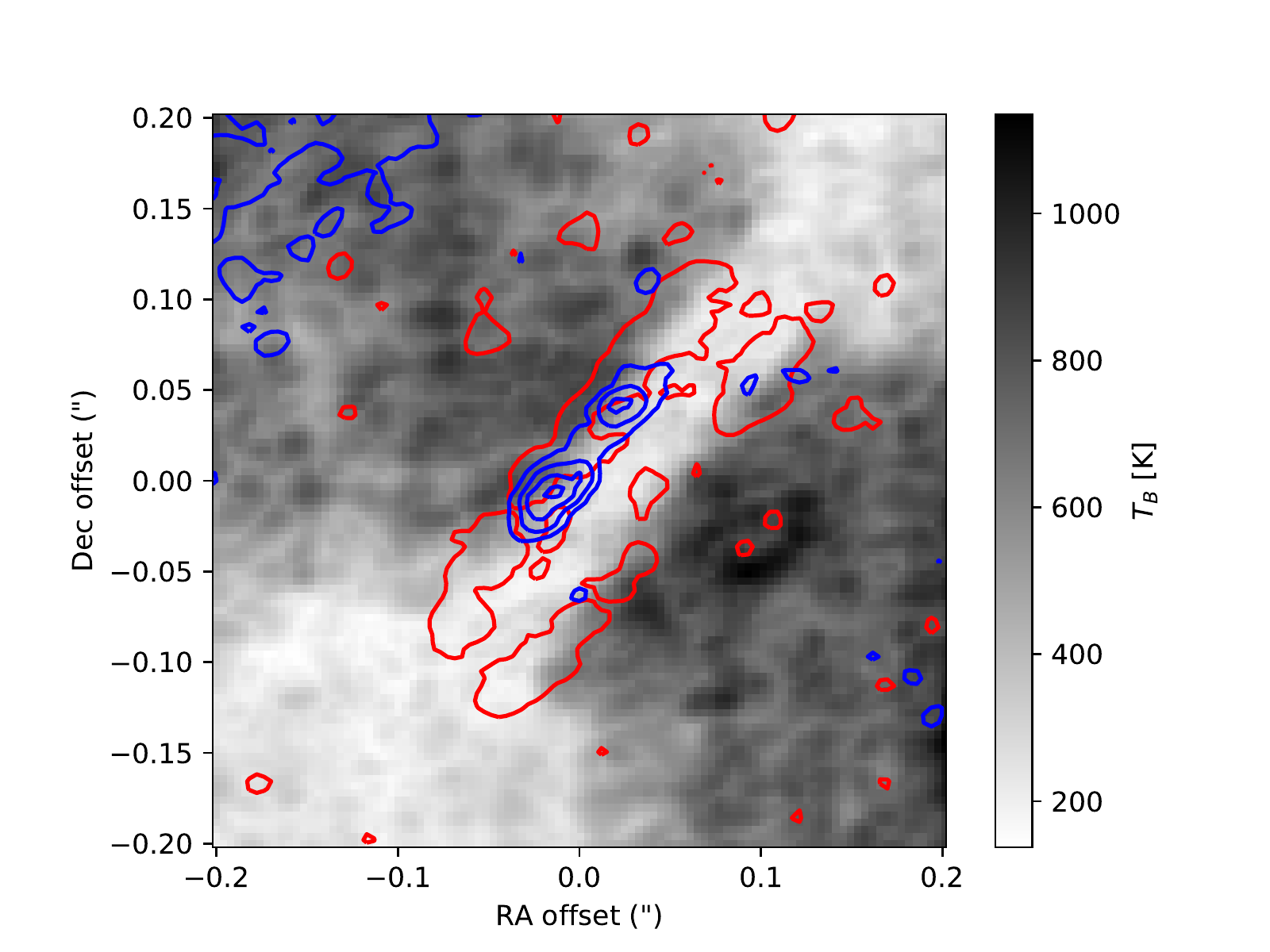}
\caption{Peak intensity map of SiO v=0 J=8-7 (gray scale), NaCl v=2 J=26-25
(red contour; 200 K) and {SiO v=5 J=8-7} (blue contours; 300, 400, 500, and 600
K).  The blue contours in the upper-left come from a blend with a different
line.
}
\label{fig:sioonnacl}
\end{figure*}

We also compare the salt emission locations to the SiO emission regions.
Figure \ref{fig:sioonnacl} shows the \mbox{SiO v=0 J=8-7} line, which is
thermally excited (it shows no sign of masing at any velocity) and traces the
outflow.  The \mbox{SiO v=5 J=8-7} line, with an upper state energy level of
8747 K, is also shown.  It clearly traces a smaller radius than the NaCl line,
but a similar height in the disk.  By contrast, the \mbox{SiO v=0 J=8-7}
emission, with $E_U=75$ K, starts above the vertical centroid of the NaCl line.
These spatial anticorrelations suggest that the difference between the
molecules' emission patterns is driven partly by excitation, since the salt
lines have upper state energy levels intermediate between the SiO v=0 and v=5
states and trace an intermediate region.

\subsection{Excitation}
The detection of vibrationally excited transitions from v=0 to v=6 at similar
brightness suggests that the vibrational excitation temperature $T_{ex,vib}$ of
the molecules is high.  However, we see sharply decreasing populations in
the higher rotational levels, indicating a much lower $T_{ex,rot}$.

\begin{figure*}[!htp]
\includegraphics[scale=1,width=3.5in]{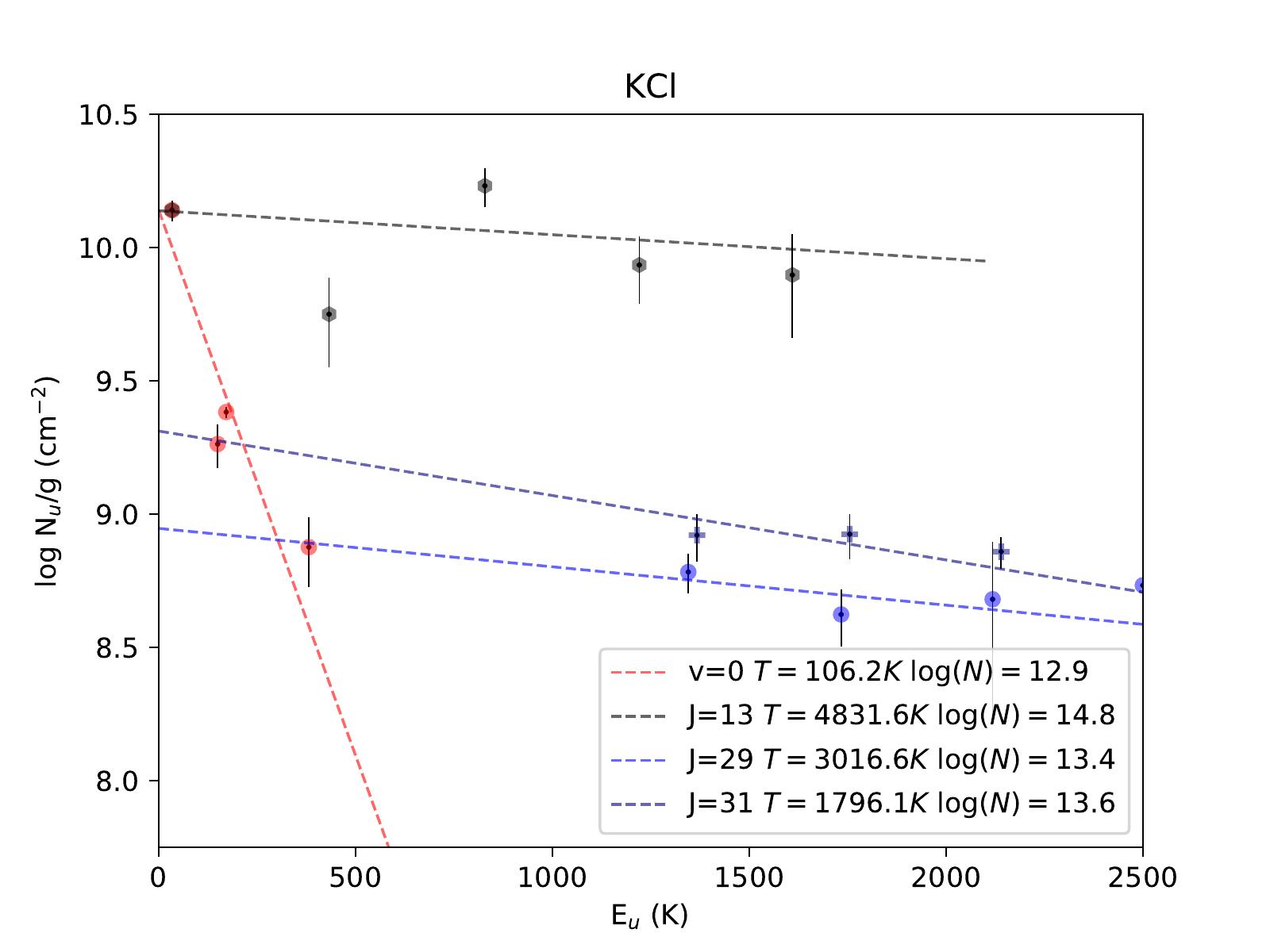}
\includegraphics[scale=1,width=3.5in]{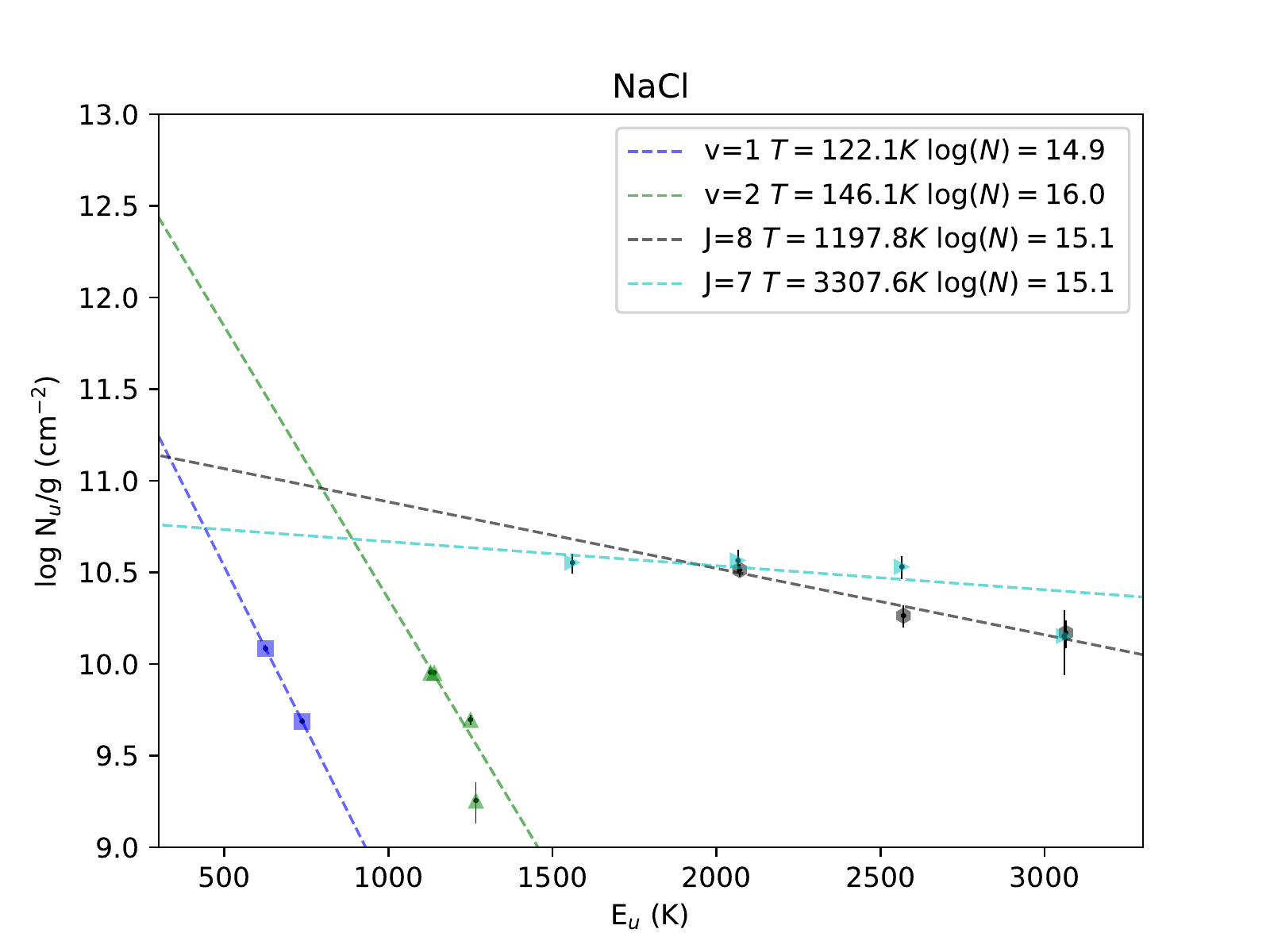}
\caption{Rotational energy diagrams for the KCl and NaCl lines.  While each
vibrational state can internally be explained reasonably well by a single
consistent rotation temperature in the range $T_{rot}\sim50-150$ K, the population
distribution between vibrational excitation states cannot.}
\label{fig:rotationdiagrams}
\end{figure*}

Within each vibrational state for which we observed several transitions, we attempt
to measure a rotational temperature.  For the KCl v=0 state, in which we have
detected 4 rotational transitions from $30 < E_U < 400$ K, the best fit
is $T_{rot}\sim105$ K (Figure
\ref{fig:rotationdiagrams}).  At such a low temperature, the populations of the
vibrationally excited states should be effectively zero.  

Similarly, for each rotational transition observed from multiple vibrational
levels, we measure a vibrational temperature.  We find values that range from
1000 to 5000 K, although these fits are subject to very large
uncertainties.
There are also statistically significant
outliers from the single-temperature model, suggesting that the level
populations may not be well-characterized by a single temperature.

Below, we explore several mechanisms for the observed line excitation:

\par{\textbf{(1) Collisional excitation by \hh.}} 
\citet{Quintana-Lacaci2016a} calculated collision rates for NaCl with He and
provided a Fortran code to compute those rates.  Typical collision rate
coefficients for \hbox{$\Delta$J=1} or \hbox{$\Delta$J=2} transitions are
$10^{-10}$~cm$^3$ \pers.  Making the usual assumption that the collision rates
with \hh are similar to those for He, we used RADEX \citep{van-der-Tak2007a} to
examine the excitation \footnote{The RADEX-compatible molecular data file
incorporating these coefficients is available from
\url{https://github.com/keflavich/SaltyDisk/blob/\githash/nacl.dat}.
\referee{We used the \texttt{pyradex} wrapper of RADEX for these
calculations (\url{https://github.com/keflavich/pyradex/}).}
}.

The NaCl transitions we observe have  Einstein A values in the range
$A\sim10^{-3}-10^{-2}$ s$^{-1}$, while the vibrationally excited states have
infrared transitions with rates $A\sim1$ s$^{-1}$
\cite{Barton2014a,Cabezas2016a}.  The collision rates into the $v=0$ states are
$C\sim10^{-11}$ cm$^3$ s$^{-1}$ (insensitive to temperature), while collision
rates into $v>0$ states from $v=0$ states are lower, $C\sim10^{-13}-10^{-12}$
cm$^3$ s$^{-1}$ at $T=1000$ K, but only $C < 10^{-15}$  cm$^3$ s$^{-1}$ at
$T=100$ K \cite{Quintana-Lacaci2016a}.  Thus, extremely high gas temperatures
and densities would be required to collisionally excite a significant
population of the molecules into vibrational states; collisional excitation
alone also cannot explain different vibrational states being observed at a
similar brightness level.

\par{\textbf{(2) Collisional excitation by electrons.}} 
Because NaCl and KCl are extremely polar molecules, with dipole moments of 9.1
and 10.3 Debye, respectively \citep{Barton2014a}, collisions with electrons
could also be important.  Using the approximations from \citet{Dickinson1975a},
we find that collision rates between both NaCl and KCl and electrons in the
J=5-50 v=0 states are $C\approx10^{-5}$ cm$^{3}$ \pers.  Thus, electrons are
$10^5$ times as efficient at exciting these molecules than \hh. In typical
low-mass disks, ionization fractions $X_e \geq 10^{-4}$ occur in the upper
layers of the disk, where some UV radiation is able to ionize H and C
\citep{Bergin2007a}.  Collisions with electrons would dominate the excitation
of the salt lines in such regions.  However, deeper into the disks only X-rays
and cosmic rays can penetrate, resulting in fractional ionizations $X_e <
10^{-6}$, too small for electron collisions to be relevant. 

The low rotational temperatures that we infer seem to rule out both \hh and
electron collisional excitation of excited vibrational states, since a
collision rate high enough to excite these vibrational levels would quickly
thermalize their rotational ladders at high temperatures.

\par{\textbf{(3) Infrared excitation.}} 
The observed emission from vibrationally excited states suggests the
presence of a significant radiation field at 25-45 \um, which covers the range
of $\Delta v=1$ rovibrational transitions with $v\leq6$ (for NaCl, the range is
25-35 \um, for KCl, it is 35-45 \um).  Since the selection rules for these
transitions require that $\Delta v=1$ \footnote{Einstein A-values for the 
$\Delta v=2$ transitions are $\sim100$ times smaller than for the $\Delta v=1$
transitions, for which $A\sim1$ \pers.}, the radiation density must be high
enough to maintain a large population in the $v=1,
2, 3, 4, 5$ states such that some fraction of photons are able to excite
molecules to still higher states.  Such a strong radiation field would
be expected to thermalize the rotational ladders within each vibrational
state to the same high temperature, which is not observed.

We experimented with several RADEX models with varying backgrounds.  The
fiducial system, with a half-sky-filling $T_{bg}=200$ K (representing the
optically thick disk) and a dilute 4000 K radiation field $I_{\nu} = (100
\mathrm{R_\odot})^2 / (30 \mathrm{au})^2 B_\nu(4000 \mathrm{K})$ (representing
the star) reasonably predicts the $v=1$ and $v=0$ intensities, but
substantially underpredicts the $v>1$ intensities.  A stronger radiation
field in the 25-45 \um regime is required, but such a field cannot be
reasonably produced by combinations of blackbodies.  If there were a dust
emission feature in that range, such as crystalline silicates forsterite
and enstatite \citep[e.g.][]{Molster2005a}, that dominated the emission spectrum
in the far-infrared, it could potentially explain the observed vibrational
excitation patterns.  Alternatively, bright emission lines from other molecules,
such as water, in the mid-infrared might also drive the peculiar excitation.

\begin{figure*}[!htp]
\includegraphics[scale=1,width=4in]{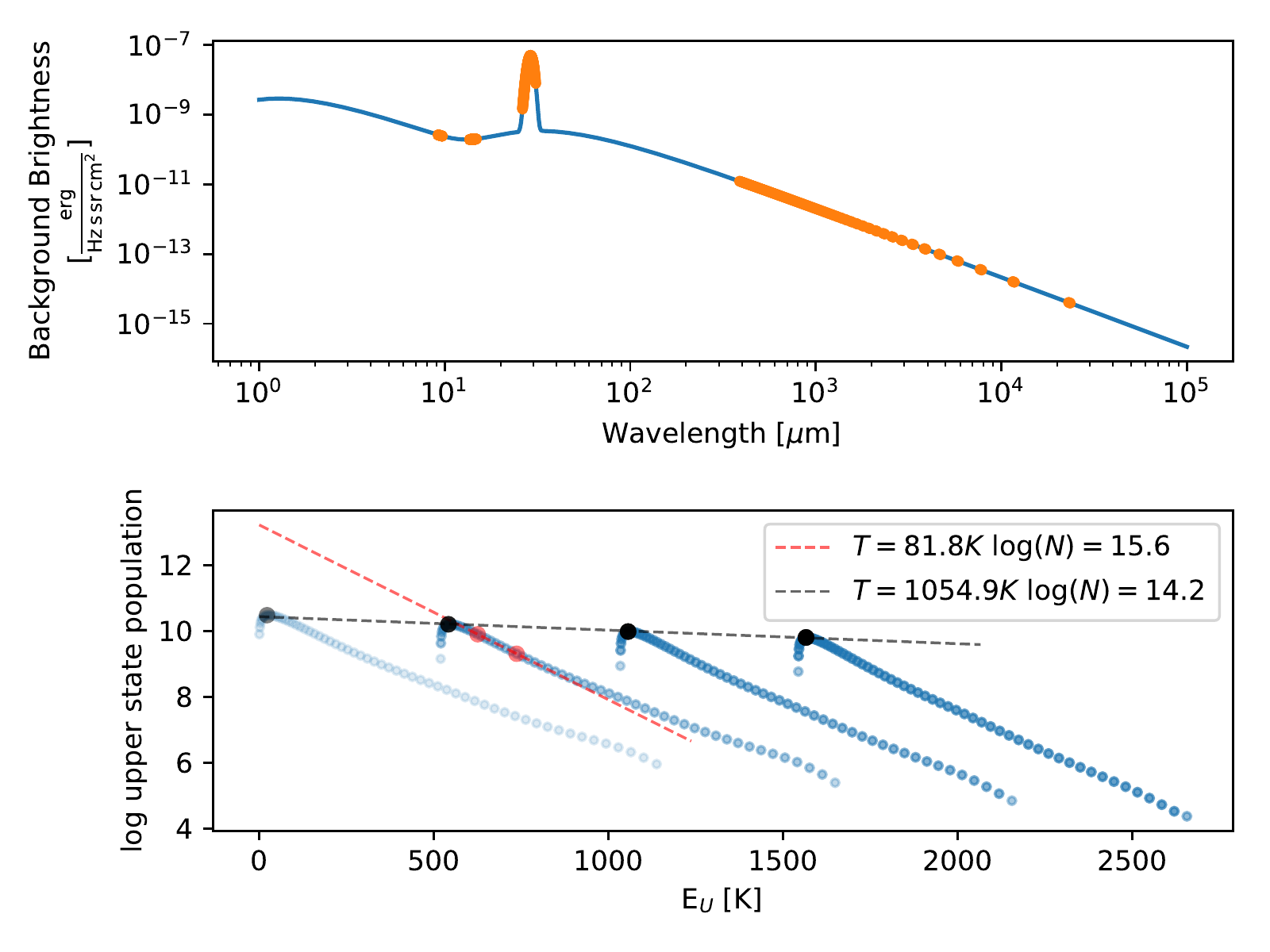}
\caption{This figure demonstrates that a bright emission feature in the mid-infrared may
explain the observed level populations.  The top panel shows the input radiation field as a blue
line, with orange dots showing the NaCl transitions.  The narrow peak at 29 \um is artificially
added, and it may represent a broad dust emission feature.
The bottom panel shows a level population diagram similar to Figure \ref{fig:rotationdiagrams}.
The models include only v=0,1,2,3 because the collisional constants are limited to these states.
The black dashed line shows a fit to the J=8 transitions of all vibrational states, yielding a high
$\sim1000$ K temperature.  The red dashed line shows a fit to the observed $v=1$ lines, obtaining a
much lower $\sim100$ K temperature.  The vertical scaling, the column density, is arbitrary\referee{, i.e., it
does not represent a fit to the data; the optical depths of the observed lines range from $\sim0.25$ to $\sim1.5$}.
}
\label{fig:naclexcitationmodel}
\end{figure*}

To demonstrate that radiative excitation in the mid-infrared is a plausible
mechanism to explain the rotation diagrams, we performed additional RADEX
modeling in which we modified the background radiation field.  We adopted a
model with a dilute $150$K blackbody representing the disk, a very dilute 4000
K blackbody representing the star, plus a strong Gaussian feature centered at
29 \um with width 1 \um.  With this radiation field, the vibrational lines 
are highly excited and exhibit a vibrational temperature significantly higher
than the rotational temperature seen within any vibrational state.
We show one example model in Figure \ref{fig:naclexcitationmodel}.
This toy model illustrates that it is possible, via radiative excitation, to obtain level
population ratios similar to those observed, but only with a peculiar radiation field\referee{; similar
experiments exploring a range of physical parameters with a blackbody radiation field could not reproduce
the different vibrational and rotational temperatures}.

\par{\textbf{(4) Ultraviolet excitation.}}
The high vibrational and low rotational temperatures of the
salt lines resemble the `sawtooth' pattern observed for the rovibrational
populations of \hh toward the Orion bar photodissociation region.  This
pattern is explained by ultraviolet excitation into higher electronic states,
followed by decay back to excited vibrational levels in the ground electronic
state \citep{Kaplan2017a}.  \hh rotational populations in the ground vibrational
state are thermalized by collisions; UV excitation transposes this pattern into
higher vibrational states.

Unfortunately, there are several problems applying this model to the salt lines.
First, we find an abrupt decline in the populations of $v > 6$ vibrational states,
with weak or ambiguous detections of NaCl in the v=7 and v=8 levels, and
nondetections for $v > 9$.  Unless there is
some selection effect in the electronic de-excitation disfavoring higher vibrational
states, it is difficult to explain this cutoff.  
Second, the lowest excited electronic energy levels for both NaCl and KCl are about
5 eV above the ground state, very close to the photodissociation threshold for these
molecules \citep{Zeiri1983, Silver1986}, which
may therefore be dissociated rather than excited by UV photons.
Third, there may not be any UV radiation in the outer disk where these lines
are detected.  
While UV
emission may be produced in shocks in the outflow, and $\sim5$ eV photons
could be produced by the central $\sim4000$ K photosphere \citep{Testi2010a},
it is unclear whether it could penetrate into the disk where the salt emission
is observed.  Even a light haze of dust in the upper layers of the disk would
likely shield the molecules from such illumination.

\par{\textbf{(5) Dust opacity.}}
\referee{
Finally, we consider the possibility that rotational temperatures,
which are based on measurements of spectral lines over a wide frequency
range, are underestimated because of greater dust extinction at high
frequencies.  Vibrational temperatures should be little affected by
extinction because they are derived from lines in the same ALMA band that
suffer similar dust attenuation.}

\referee{
Comparison of the observed line and continuum ratios across the bands suggests
that dust extinction is not the dominant effect governing the rotational
temperature.
The KCl v=0 lines, which are the only vibrational state detected in all three observed
bands, have similar peak brightnesses at $\sim100$ and $\sim345$ GHz
($T_{B,max}(\textrm{KCl v=0 13-12}) = 15.3\pm0.9$ K and $T_{B,max}(\textrm{KCl
v=0 45-44}) = 9.7\pm1.8$ K).  
If the rotational excitation temperature $T_{ex}$ were comparable
with the observed vibrational excitation temperatures $T_{vib} \sim 1000$ K, 
the 45-44 transition would be 9 times as bright as the 13-12 transition.
If we assume a dust opacity index $\beta=2$,
which is conservative in that it provides the greatest difference in extinction
between the two frequencies, the dust opacities required to produce the observed line ratio
are $\tau_{345 \textrm{GHz}} = 2.8$ and $\tau_{100 \textrm{GHz}} = 0.2$.  These optical
depths  would result in a substantially ($\approx4\times$) higher 
dust brightness at 0.85 \um than at 3 mm, which is not observed.
}

\begin{figure*}[!htp]
\includegraphics[scale=1,width=7in]{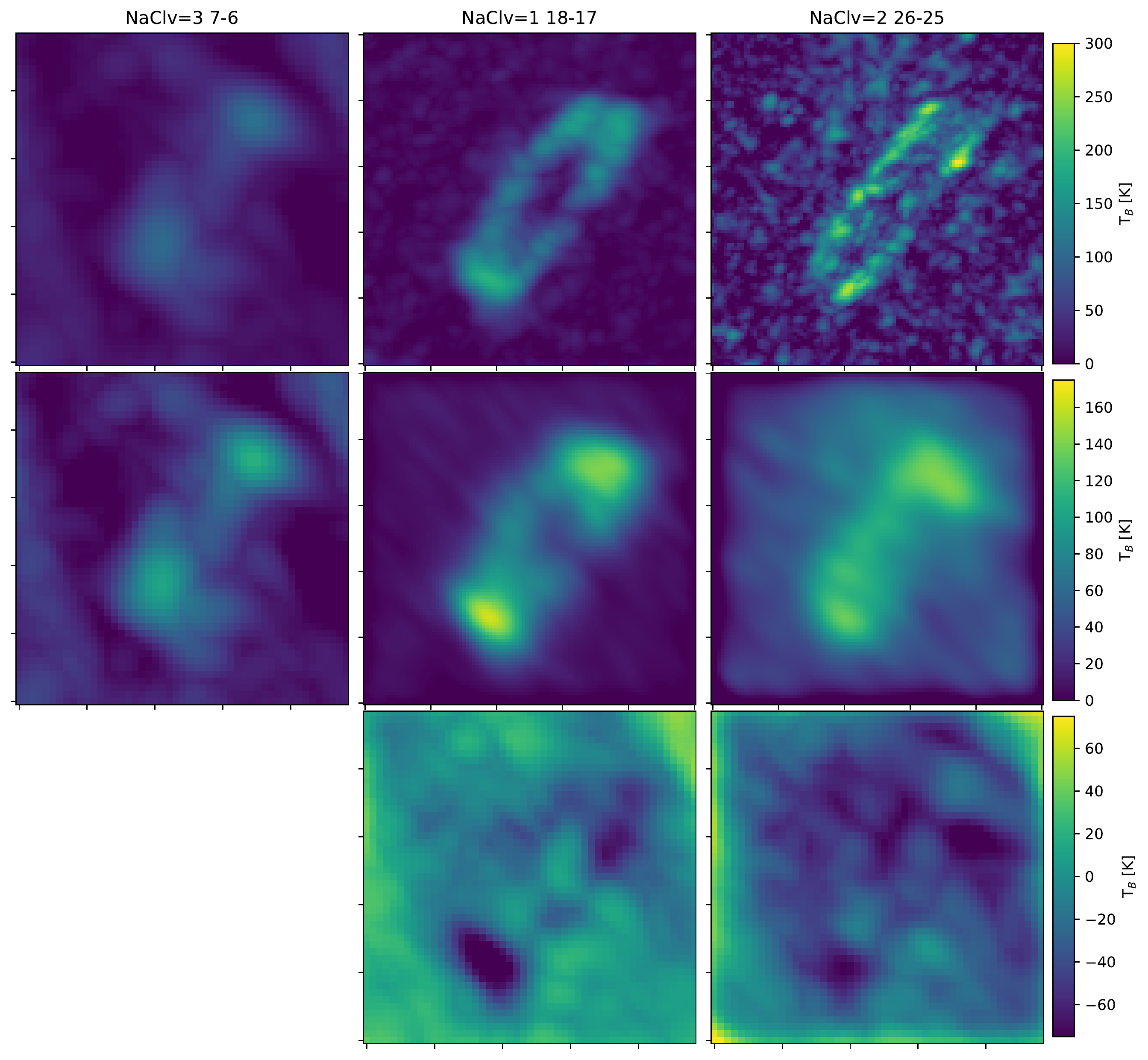}
\caption{Comparison of three NaCl emission lines in different wavebands viewed
at the same spatial resolution.  The columns show, from left to right, NaCl v=3
J=7-6, NaCl v=1 J=18-17, and NaCl v=2 J=26-25, as in Figure \ref{fig:spatial}.
The top row shows these images in their native resolution, the middle shows
them convolved to the resolution of the 3 mm band (note that the 3 mm band
image is unchanged, but is repeated because the color scale has changed), and
the bottom row shows the difference images $T_{B,3 \mathrm{mm}} - T_{B, 1.4
\mathrm{mm}}$ and $T_{B,3 \mathrm{mm}} - T_{B, 0.87 \mathrm{mm}}$.  }

\label{fig:resolutioncomparison}
\end{figure*}

\referee{
Dust obscuration on sub-beam scales could preferentially block our view of some
of the high frequency salt emission region, which would bias our measurements
toward cooler rotational temperatures.  However, the morphology of the line
emission does not change with frequency, suggesting that this effect is
negligible.
Figure \ref{fig:resolutioncomparison} shows the salt line images in each
frequency band convolved to the same resolution, confirming that the morphological
differences are minimal.
}

\bigskip

We conclude that there is no simple model that explains the observed pattern of
low rotational and high vibrational temperature.

\subsection{What is producing gas-phase salts?}

Sodium and potassium are rarely observed in the dense molecular interstellar
medium.  They are usually assumed to be rapidly incorporated into dust grains
after being ejected from dying stars \citep[e.g.,][]{Milam2007a}.  Since we
observe NaCl and KCl in the atmosphere of the disk, it is clear that
there is a zone where either dust has not  formed  (which is not very likely
in a protostellar system)
or
where dust is destroyed and returned to the gas phase.  We argue that the dust
must be destroyed almost immediately as it is launched into the disk-driven
outflow \citep{Hirota2017b}. We evaluate three scenarios for salt production:
sputtering off of dust by high-energy particles, gas-phase formation from
atomic gas, and thermal desorption.

\textbf{(1) Sputtering:}
In the first scenario, NaCl and KCl molecules are sputtered from the grains.
The most plausible means to achieve the energies required to sputter grains is
with strong shocks \citep{Schilke1997a,Decin2016a}.  In \sourcei, it seems
unlikely that there would be very strong shocks in the disk, but there are
high-velocity shocks throughout the outflow.  Bright SiO emission is observed in
the inner part of the outflow (i.e., within $<100$ AU of \sourcei), suggesting
that grains are efficiently destroyed there.  Without knowing the grain
structure, however, it is unclear whether high-energy particles capable of
destroying the grain cores are present.

Since the salts are detected close to but above the disk midplane, the
grains must be sputtered very rapidly after being launched from the disk surface.
The lack of salt and SiO emission in the disk midplane may be because these
molecules are not in the gas phase at all in the midplane.  Alternatively,
radiative transfer can explain their non-detection, i.e., if the continuum
background is the same intensity as the emission lines.

\referee{
In this model, both SiO and the salts are in the gas phase above the disk
midplane. 
However, while low-J SiO lines are seen throughout the outflow at very high
elevations above the disk, the salt lines are limited to $\lesssim 30$ AU from
the midplane.  Even the low-J salt lines, which can be excited at the same
modest densities required to excite SiO  ($n\sim10^4$ to $10^6$ \percc), are
not observed at high elevations.  Excitation therefore cannot
explain the morphological difference between SiO and the salts, and instead we
infer that they are chemically distinct.  Possible explanations are that the salts
deplete back onto grains as they are entrained in the outflow, they react to
form other molecules, or they dissociate into atomic form while SiO remains
molecular.
}

\textbf{(2) Gas-phase formation:}
Gas-phase formation of NaCl and KCl is possible from ionized Na and K.
Potassium and sodium have similar first ionization potentials at 4.34 and 5.14
eV, respectively.  These atoms would be nearly fully ionized between 1500-2000
K at a density of $10^6$ \percc (aluminum has a higher first ionization
potential of 5.99 eV, and so is ionized at a few hundred K higher temperatures
than sodium).  These ions react with HCl to form NaCl and KCl.  However, at
present, only the inner $<20$ AU region, which is currently unresolved, clearly
reaches such high temperatures\footnote{We also know from RRL studies that
there is little to no ionized gas with $T\gtrsim10^4$ K around \sourcei
\citep{Plambeck2016a,Baez-Rubio2018a}}.  

For gas-phase formation to be a significant process, then, either material in
that hot inner region is being transported outward, which seems unlikely since
there is a bulk outflow lifting material off of the disk, or the disk was
previously substantially warmer.  Under the dynamical formation scenario of the
disk, in which it is the remnant of a previously larger disk from the
\sourcei-BN interaction \citep{Bally2017a,Luhman2017a,Kim2018f},  
energy was released as the disk re-settled into its current apparently smooth
state, which could have resulted in much higher gas temperatures over the last
few hundred years.  This scenario would imply that the observable state of NaCl
and KCl is very short lived, as these molecules are in the process of depleting
onto grains.  While plausible, there are many processes that need to coincide
perfectly in this scenario (disk temperature history, gas density, gas-phase
formation rates), so we favor the dust destruction hypotheses.

We note that, if the formation rates of NaCl and KCl are very high, it is
possible that the vibrational excitation observed is from the chemical energy
of formation.  However, because of the short expected lifetimes of the
vibrationally excited states, we regard this explanation for the excitation as
very unlikely.

\textbf{(3) Thermal desorption from grain surfaces}
\citet{Decin2016a} presented a model for the gas-phase abundance of
NaCl in which it is thermally desorbed from grains at grain temperatures
between 100 and 300 K, depending on the unknown binding energy of NaCl with
the grain surface.  Since we expect grain temperatures $T>150$ K in the
NaCl emission region (see Section \ref{sec:wherefrom}), this model hints
that simple grain heating is enough to explain the high gas-phase abundances
of NaCl observed.  Since NaCl is not observed within $r<30 AU$, though,
it must be dissociated at smaller radii, which most likely requires
a substantial UV field within that radius in the wavelength range
$912 \mathrm{\AA} < \lambda < 3000 \mathrm{\AA}$ \citep{Silver1986a}.

This model is plausible, though it relies on unknown physical parameters.
Additionally, while a 4000 K stellar atmosphere \citep{Testi2010a} produces
non-negligible radiation short of 3000 \AA, it is unclear whether that
radiation can propagate to the part of the disk where the salts are observed,
since small dust grains may pervade the system.

\subsection{Other salts}
While NaCl and KCl are clearly detected here, there are also several
transitions of AlCl and AlF in our observational bands that were not detected.
\citet{Cernicharo1987a} detected these transitions at comparable brightness to
NaCl and KCl in IRC+10216.  The lack of AlCl may be because \sourcei's disk is
oxygen-rich, and aluminum is locked into AlOH, as suggested in the
\citet{Cherchneff2012a} model.  Indeed, the tentative detection of AlO
in our spectra supports this conclusion.

\subsection{Isotopologue abundances}
With several transitions of each of the observed isotopologues, we can in
principle measure the relative abundance of the various isotopes.  However,
because of the uncertainties in the excitation above, these measurements should
be taken with a grain of salt.

The $^{37}$Cl abundance can be measured by comparing the intensity of the NaCl
and Na$^{37}$Cl v=3 J=7-6, v=4 J=8-7 and J=7-6, and v=5 J=7-6 lines
\referee{assuming they are optically thin}.  The ratio $^{35}$/$^{37}$Cl is
$r=1.4\pm0.4$.  The KCl v=0 J=45-44 line gives a $^{35}/^{37}$Cl ratio $r=0.8$.
\citet{Agundez2012a} measured this ratio to be $r=2.9\pm0.3$ in IRC+10216, and
\citet{Highberger2003a} reported a ratio of 2.1 $\pm$ 0.8 in CRL 2688.  Our
measurement is somewhat lower, but should be regarded as consistent at least
until we can obtain more accurate column density measurements.

We only have one commonly observed $^{41}$KCl / $^{39}$KCl line, v=0 J=13-12,
which has a ratio $^{39}/^{41}$ $r=2.4$.

\subsection{Future prospects}
The detection of these refractory species in the gas phase in the atmosphere
of \sourcei's disk hints that other refractory molecules may be present, 
which could enable their first detection in the ISM (e.g., such rare species as
FeO and FeS), and may enable direct measurements of the metallicity in
star-forming gas.

The observed NaCl and KCl transitions uniquely trace the disk, unlike most other
molecules observed in the ISM that are bright and abundant in hot cores like
Sgr B2 \citep{Nummelin1998a,Belloche2013a}.  These lines therefore potentially
represent tracers that can uniquely trace high-mass disks, making them extremely
valuable probes of early-stage high-mass star formation\referee{, at least at
distances where the salt-emitting regions can be resolved}.
However, it remains unclear whether \sourcei is a unique source or is
representative of the class of high-mass protostars with disks; if the latter,
these lines and potentially many others can be used to probe the radiation
environment of extremely embedded high-mass protostars.  An obvious next step
is to search for these lines in other resolved disks around high-mass
protostars \citep[e.g.,  HH80/81;][]{Girart2017a}.  These species are
particularly interesting for future millimeter and centimeter wavelength
observatories like the ngVLA and SKA, since they have lower-J transitions into
the few centimeter range, wavelengths at which the dust is very likely
to be optically thin even in the inner few AU of a disk.

The detection of these species with substantial population in vibrationally
excited states highlights the need for laboratory work to measure such
transitions in known astronomical molecular species. Indeed, prior work on
vibrational excitation in modeling and assigned vibrationally excited ethyl
cyanide in ALMA observations of Orion-KL revealed that even when catalogs for
these states are available, they are often incomplete \citep{Fortman2012a}.

\begin{table*}[htp]
\centering
\caption{Parameters of NaCl lines obtained with Gaussian fits}
\begin{tabular}{ccccccc}
\label{tab:NaCl_salt_lines}
 J$_u$ & J$_l$ & Frequency & Velocity & Amplitude & $\int T_A dv$ & E$_U$ \\
  &  & $\mathrm{GHz}$ & $\mathrm{km\,s^{-1}}$ & $\mathrm{K}$ & $\mathrm{K\,km\,s^{-1}}$ & $\mathrm{K}$ \\
\hline
&\vspace{-0.75em}\\
\multicolumn{7}{c}{$v = 1$} \\
\vspace{-0.75em}\\
 26 & 25 & 335.50656 & 5.5 (0.1) & 29.1 (0.8) & 339.9 (5.4) & 737.2 \\
 18 & 17 & 232.50998 & 5.4 (0.2) & 23.9 (0.7) & 396.6 (6.7) & 625.7 \\
&\vspace{-0.75em}\\
\multicolumn{7}{c}{$v = 2$} \\
\vspace{-0.75em}\\
 17 & 16 & 217.98023 & 5.2 (0.1) & 19.3 (0.4) & 267.8 (3.6) & 1128.4 \\
 18 & 17 & 230.77917 & 5.6 (0.1) & 19.3 (0.4) & 297.7 (3.7) & 1139.5 \\
 26 & 25 & 333.00729 & 7.4 (0.3) & 21.7 (0.9) & 352.2 (8.9) & 1250.2 \\
 27 & 26 & 345.76204 & 0.2 (0.9) & 11.5 (1.9) & 136.4 (13.8) & 1266.8 \\
&\vspace{-0.75em}\\
\multicolumn{7}{c}{$v = 3$} \\
\vspace{-0.75em}\\
 7 & 6 & 89.15011 & 3.8 (0.5) & 14.1 (1.1) & 183.1 (9.0) & 1561.0 \\
&\vspace{-0.75em}\\
\multicolumn{7}{c}{$v = 4$} \\
\vspace{-0.75em}\\
 7 & 6 & 88.48549 & 3.8 (0.7) & 11.4 (1.0) & 185.4 (10.5) & 2065.5 \\
 8 & 7 & 101.12183 & 3.1 (0.3) & 15.5 (0.8) & 214.1 (7.0) & 2070.4 \\
&\vspace{-0.75em}\\
\multicolumn{7}{c}{$v = 5$} \\
\vspace{-0.75em}\\
 8 & 7 & 100.36722 & 3.7 (0.5) & 10.0 (0.9) & 122.3 (6.6) & 2570.0 \\
 7 & 6 & 87.82519 & 4.5 (0.6) & 11.6 (1.1) & 173.8 (9.9) & 2565.2 \\
&\vspace{-0.75em}\\
\multicolumn{7}{c}{$v = 6$} \\
\vspace{-0.75em}\\
 8 & 7 & 99.61753 & 4.4 (0.7) & 7.7 (0.9) & 100.0 (6.8) & 3064.8 \\
 7 & 6 & 87.16921 & 3.4 (1.2) & 7.2 (1.8) & 73.9 (11.4) & 3060.0 \\
\hline
\end{tabular}

\par 
\end{table*}
\begin{table*}[htp]
\centering
\caption{Parameters of Na$^{37}$Cl lines obtained with Gaussian fits}
\begin{tabular}{ccccccc}
\label{tab:Na37Cl_salt_lines}
 J$_u$ & J$_l$ & Frequency & Velocity & Amplitude & $\int T_A dv$ & E$_U$ \\
  &  & $\mathrm{GHz}$ & $\mathrm{km\,s^{-1}}$ & $\mathrm{K}$ & $\mathrm{K\,km\,s^{-1}}$ & $\mathrm{K}$ \\
\hline
&\vspace{-0.75em}\\
\multicolumn{7}{c}{$v = 0$} \\
\vspace{-0.75em}\\
 7 & 6 & 89.22011 & 4.3 (0.4) & 18.2 (1.2) & 230.2 (8.9) & 17.1 \\
 18 & 17 & 229.24605 & 4.8 (0.1) & 21.4 (0.4) & 251.5 (3.2) & 104.6 \\
&\vspace{-0.75em}\\
\multicolumn{7}{c}{$v = 1$} \\
\vspace{-0.75em}\\
 17 & 16 & 214.93871 & 4.8 (0.4) & 18.7 (1.2) & 217.3 (8.8) & 607.0 \\
 8 & 7 & 101.21188 & 4.6 (0.3) & 19.7 (0.8) & 281.1 (7.1) & 536.0 \\
&\vspace{-0.75em}\\
\multicolumn{7}{c}{$v = 2$} \\
\vspace{-0.75em}\\
 8 & 7 & 100.46695 & 5.3 (0.4) & 15.6 (0.7) & 307.2 (9.0) & 1045.0 \\
 7 & 6 & 87.91232 & 4.3 (0.6) & 13.1 (1.1) & 205.3 (10.2) & 1040.1 \\
&\vspace{-0.75em}\\
\multicolumn{7}{c}{$v = 3$} \\
\vspace{-0.75em}\\
 7 & 6 & 87.26464 & 2.9 (1.0) & 9.0 (1.7) & 102.0 (11.9) & 1544.3 \\
 8 & 7 & 99.72675 & 3.7 (0.4) & 14.0 (0.8) & 203.9 (7.3) & 1549.1 \\
&\vspace{-0.75em}\\
\multicolumn{7}{c}{$v = 4$} \\
\vspace{-0.75em}\\
 7 & 6 & 86.62109 & -2.1 (1.1) & 13.0 (1.4) & 265.3 (19.6) & 2043.6 \\
 8 & 7 & 98.99127 & 4.6 (1.1) & 8.2 (1.9) & 86.3 (11.9) & 2048.4 \\
 27 & 26 & 333.45906 & 5.5 (0.5) & 10.9 (1.0) & 130.9 (7.3) & 2251.3 \\
&\vspace{-0.75em}\\
\multicolumn{7}{c}{$v = 5$} \\
\vspace{-0.75em}\\
 19 & 18 & 233.16920 & 5.2 (0.5) & 7.4 (0.8) & 72.8 (5.0) & 2633.6 \\
 7 & 6 & 85.98167 & 5.2 (1.2) & 8.1 (1.7) & 99.9 (12.6) & 2538.2 \\
\hline
\end{tabular}

\par 
\end{table*}
\begin{table*}[htp]
\centering
\caption{Parameters of KCl lines obtained with Gaussian fits}
\begin{tabular}{ccccccc}
\label{tab:KCl_salt_lines}
 J$_u$ & J$_l$ & Frequency & Velocity & Amplitude & $\int T_A dv$ & E$_U$ \\
  &  & $\mathrm{GHz}$ & $\mathrm{km\,s^{-1}}$ & $\mathrm{K}$ & $\mathrm{K\,km\,s^{-1}}$ & $\mathrm{K}$ \\
\hline
&\vspace{-0.75em}\\
\multicolumn{7}{c}{$v = 0$} \\
\vspace{-0.75em}\\
 28 & 27 & 215.00828 & 5.5 (0.6) & 10.4 (1.3) & 113.1 (8.4) & 149.7 \\
 30 & 29 & 230.32064 & 4.6 (0.2) & 14.4 (0.4) & 171.2 (3.2) & 171.4 \\
 45 & 44 & 344.82061 & 2.5 (1.1) & 9.7 (1.8) & 120.0 (13.9) & 381.2 \\
 13 & 12 & 99.92952 & 4.0 (0.3) & 15.3 (0.9) & 183.8 (6.5) & 33.6 \\
&\vspace{-0.75em}\\
\multicolumn{7}{c}{$v = 1$} \\
\vspace{-0.75em}\\
 13 & 12 & 99.31663 & 3.9 (1.0) & 8.3 (2.0) & 75.2 (11.0) & 432.6 \\
 44 & 43 & 335.13396 & 3.3 (0.3) & 14.1 (0.8) & 143.3 (5.0) & 761.7 \\
&\vspace{-0.75em}\\
\multicolumn{7}{c}{$v = 2$} \\
\vspace{-0.75em}\\
 44 & 43 & 333.06770 & 3.9 (0.5) & 11.4 (1.0) & 135.6 (7.3) & 1155.3 \\
 13 & 12 & 98.70595 & 2.0 (0.8) & 14.2 (1.5) & 229.2 (15.2) & 828.3 \\
&\vspace{-0.75em}\\
\multicolumn{7}{c}{$v = 3$} \\
\vspace{-0.75em}\\
 31 & 30 & 233.60570 & 4.8 (0.7) & 6.0 (0.8) & 64.1 (5.2) & 1367.1 \\
 13 & 12 & 98.09753 & 6.1 (1.1) & 9.2 (1.7) & 116.3 (13.2) & 1220.6 \\
 29 & 28 & 218.57971 & 4.4 (0.4) & 5.0 (0.6) & 40.8 (2.8) & 1345.1 \\
&\vspace{-0.75em}\\
\multicolumn{7}{c}{$v = 4$} \\
\vspace{-0.75em}\\
 31 & 30 & 232.16185 & 4.6 (0.6) & 6.6 (0.8) & 65.0 (5.0) & 1755.1 \\
 13 & 12 & 97.49133 & 8.0 (2.1) & 6.1 (1.5) & 107.3 (18.0) & 1609.5 \\
 29 & 28 & 217.22891 & -0.3 (0.6) & 3.5 (0.6) & 28.4 (2.8) & 1733.2 \\
&\vspace{-0.75em}\\
\multicolumn{7}{c}{$v = 5$} \\
\vspace{-0.75em}\\
 45 & 44 & 334.29930 & 2.1 (0.3) & 42.9 (0.5) & 1550.0 (21.5) & 2332.2 \\
 31 & 30 & 230.72399 & 5.3 (0.4) & 5.2 (0.5) & 56.2 (3.1) & 2139.8 \\
 29 & 28 & 215.88373 & 6.3 (2.1) & 3.0 (1.3) & 32.6 (8.4) & 2118.0 \\
&\vspace{-0.75em}\\
\multicolumn{7}{c}{$v = 6$} \\
\vspace{-0.75em}\\
 31 & 30 & 229.29217 & 6.0 (0.5) & 4.1 (0.5) & 32.6 (2.6) & 2521.2 \\
 47 & 46 & 346.87489 & 0.7 (1.1) & 4.4 (0.8) & 61.1 (6.8) & 2745.3 \\
 29 & 28 & 214.54412 & 6.3 (1.9) & 3.4 (1.3) & 37.0 (8.5) & 2499.6 \\
&\vspace{-0.75em}\\
\multicolumn{7}{c}{$v = 7$} \\
\vspace{-0.75em}\\
 47 & 46 & 344.71476 & 4.7 (3.1) & 5.1 (1.4) & 118.3 (23.7) & 3122.1 \\
\hline
\end{tabular}

\par 
\end{table*}
\begin{table*}[htp]
\centering
\caption{Parameters of K$^{37}$Cl lines obtained with Gaussian fits}
\begin{tabular}{ccccccc}
\label{tab:K37Cl_salt_lines}
 J$_u$ & J$_l$ & Frequency & Velocity & Amplitude & $\int T_A dv$ & E$_U$ \\
  &  & $\mathrm{GHz}$ & $\mathrm{km\,s^{-1}}$ & $\mathrm{K}$ & $\mathrm{K\,km\,s^{-1}}$ & $\mathrm{K}$ \\
\hline
&\vspace{-0.75em}\\
\multicolumn{7}{c}{$v = 0$} \\
\vspace{-0.75em}\\
 45 & 44 & 335.05072 & 7.0 (0.4) & 12.6 (0.7) & 182.0 (6.1) & 370.4 \\
&\vspace{-0.75em}\\
\multicolumn{7}{c}{$v = 1$} \\
\vspace{-0.75em}\\
 47 & 46 & 347.71265 & 4.6 (0.5) & 7.4 (1.0) & 63.7 (5.1) & 794.8 \\
 31 & 30 & 229.81880 & 5.5 (0.3) & 7.2 (0.4) & 86.9 (3.2) & 570.2 \\
&\vspace{-0.75em}\\
\multicolumn{7}{c}{$v = 2$} \\
\vspace{-0.75em}\\
 12 & 11 & 88.54307 & 5.2 (1.5) & 4.3 (1.2) & 50.1 (8.5) & 811.5 \\
&\vspace{-0.75em}\\
\multicolumn{7}{c}{$v = 3$} \\
\vspace{-0.75em}\\
 12 & 11 & 88.00523 & 4.4 (1.4) & 5.6 (1.0) & 90.3 (10.4) & 1198.3 \\
&\vspace{-0.75em}\\
\multicolumn{7}{c}{$v = 4$} \\
\vspace{-0.75em}\\
 32 & 31 & 232.90755 & 5.0 (1.1) & 4.0 (0.7) & 51.7 (5.7) & 1739.2 \\
&\vspace{-0.75em}\\
\multicolumn{7}{c}{$v = 6$} \\
\vspace{-0.75em}\\
 32 & 31 & 230.07072 & 3.9 (1.3) & 1.9 (0.4) & 25.0 (3.4) & 2494.7 \\
 30 & 29 & 215.73679 & 5.9 (2.8) & 1.9 (1.5) & 13.9 (7.0) & 2473.0 \\
\hline
\end{tabular}

\par 
\end{table*}
\begin{table*}[htp]
\centering
\caption{Parameters of $^{41}$KCl lines obtained with Gaussian fits}
\begin{tabular}{ccccccc}
\label{tab:41KCl_salt_lines}
 J$_u$ & J$_l$ & Frequency & Velocity & Amplitude & $\int T_A dv$ & E$_U$ \\
  &  & $\mathrm{GHz}$ & $\mathrm{km\,s^{-1}}$ & $\mathrm{K}$ & $\mathrm{K\,km\,s^{-1}}$ & $\mathrm{K}$ \\
\hline
&\vspace{-0.75em}\\
\multicolumn{7}{c}{$v = 0$} \\
\vspace{-0.75em}\\
 29 & 28 & 217.54317 & 5.5 (0.7) & 4.5 (0.4) & 72.9 (4.0) & 156.7 \\
 13 & 12 & 97.62809 & 4.7 (2.5) & 6.5 (1.3) & 162.6 (24.7) & 32.8 \\
&\vspace{-0.75em}\\
\multicolumn{7}{c}{$v = 1$} \\
\vspace{-0.75em}\\
 45 & 44 & 334.85439 & 5.1 (0.6) & 5.2 (1.0) & 36.3 (4.2) & 764.9 \\
&\vspace{-0.75em}\\
\multicolumn{7}{c}{$v = 2$} \\
\vspace{-0.75em}\\
 31 & 30 & 229.68227 & 2.7 (0.6) & 9.6 (0.3) & 319.8 (10.2) & 962.5 \\
 12 & 11 & 89.03129 & - (-) & - (-) & - (-) & 813.8 \\
\hline
\end{tabular}

\par 
\end{table*}
\begin{table*}[htp]
\centering
\caption{Parameters of $^{41}$K$^{37}$Cl lines obtained with Gaussian fits}
\begin{tabular}{ccccccc}
\label{tab:41K37Cl_salt_lines}
 J$_u$ & J$_l$ & Frequency & Velocity & Amplitude & $\int T_A dv$ & E$_U$ \\
  &  & $\mathrm{GHz}$ & $\mathrm{km\,s^{-1}}$ & $\mathrm{K}$ & $\mathrm{K\,km\,s^{-1}}$ & $\mathrm{K}$ \\
\hline
&\vspace{-0.75em}\\
\multicolumn{7}{c}{$v = 7$} \\
\vspace{-0.75em}\\
 48 & 47 & 334.32791 & -9.5 (0.0) & 24.5 (0.9) & 347.9 (8.2) & 3049.3 \\
\hline
\end{tabular}

\par 
\end{table*}

\section{Conclusions}
We have identified many transitions of NaCl, KCl, and their isotopologues in
the disk of \sourcei.  These lines trace material very near the surface of the
disk, providing a uniquely powerful probe of the disk kinematics and physical
conditions.

Despite the wide range of transitions observed, the excitation mechanism and
conditions remain uncertain.  Further observations of lower-energy vibrational
and rotational states of these molecules will help distinguish between
radiative and collisional excitation scenarios and will either provide
direct measurements of the density or the radiation field in the 30-60 AU
region around \sourcei.

The narrow vertical extent of the salt emission indicates that dust is
destroyed nearly immediately after being raised from the surface of the disk.
This morphology hints that immediate dust destruction is an integral part of
the outflow driving process.

Given the rarity of these molecules, if they can be detected elsewhere,
they will serve as a powerful and unique probe of local conditions.
They may be the best available molecular tool to find disks around high-mass
protostars and measure their kinematics.

\acknowledgements
We thank the anonymous referee for a constructive and detailed review.
The National Radio Astronomy Observatory is a facility of the National Science
Foundation operated under cooperative agreement by Associated Universities,
Inc. The Green Bank Observatory is a facility of the National Science
Foundation operated under cooperative agreement by Associated Universities,
Inc. Support for B.A.M. was provided by NASA through Hubble Fellowship grant
\#HST-HF2-51396 awarded by the Space Telescope Science Institute, which is
operated by the Association of Universities for Research in Astronomy, Inc.,
for NASA, under contract NAS5-26555. 
This paper makes use of the following ALMA data: ADS/JAO.ALMA\#2016.1.00165.S
ALMA is a partnership of ESO (representing its member states), NSF (USA) and
NINS (Japan), together with NRC (Canada), MOST and ASIAA (Taiwan), and KASI
(Republic of Korea), in cooperation with the Republic of Chile. The Joint ALMA
Observatory is operated by ESO, AUI/NRAO and NAOJ.

\software{
The software used to make this version of the paper is available from github at
\url{https://github.com/keflavich/Orion_ALMA_2016.1.00165.S}
with hash \githash
(\gitdate).
The data are available from Zenodo at 
\dataset[doi:10.5281/zenodo.1213350]{https://doi.org/10.5281/zenodo.1213350}. 
The tools used include \texttt{spectral-cube}
\citep[][and \url{https://github.com/radio-astro-tools/spectral-cube}]{Ginsburg2018SpectralCube}
and
\texttt{radio-beam} 
\citep[][and \url{https://github.com/radio-astro-tools/radio-beam}]{Ginsburg2018RadioBeam}
from the
\texttt{radio-astro-tools} package
(\url{radio-astro-tools.github.io}), \texttt{astropy}
\citep{Astropy-Collaboration2013a},
\texttt{pyradex} \url{https://github.com/keflavich/pyradex/},
\texttt{scipy} \citep{Scipy2001},
\texttt{numpy} \citep{Numpy2006a},
\texttt{matplotlib} \citep{Matplotlib},
\texttt{astroquery}
\citep[][and~\url{astroquery.readthedocs.io}]{Ginsburg2018Astroquery}
and \texttt{CASA} \citep{McMullin2007a}.
A script to produce the collision rate table file using the fortran code
provided by \citet{Quintana-Lacaci2016a} is available at
\url{https://github.com/keflavich/Orion_ALMA_2016.1.00165.S/blob/master/analysis/collision_rates_nacl.py}.
}

\end{document}